\documentclass[review]{elsarticle}

\usepackage[english]{babel} 
\usepackage[T1]{fontenc}
\usepackage{lmodern} 
\usepackage[ansinew]{inputenc}
\usepackage{amsmath}
\usepackage{xfrac}
\usepackage{tikz}
\usepackage{tikz-dimline}
\pgfplotsset{
compat=1.5,
legend image code/.code={
\draw[mark repeat=2,mark phase=2]
plot coordinates {
(0cm,0cm)
(0.15cm,0cm)        
(0.3cm,0cm)         
};
}
}
\usepackage{geometry}
\usepackage{blindtext}
\usepackage{float}
\usepackage{graphicx}
\usepackage{caption}
\usepackage{hyperref}
\usepackage{pgfplots}
\usepackage{array}
\usepackage{siunitx}
\usepackage{multirow}
\usepgfplotslibrary{groupplots,dateplot}
\usetikzlibrary{patterns,shapes.arrows,calc,external}
\tikzset{>=latex}
\pgfplotsset{compat=newest}

\biboptions{authoryear}

\begin{document}

\title{Inferring incompressible two-phase flow fields from the interface motion using physics-informed neural networks}

\author{Aaron B. Buhendwa\corref{cor}}
\ead{aaron.buhendwa@tum.de}
\cortext[cor]{Corresponding author}

\author{Stefan Adami\corref{}}
\ead{stefan.adami@tum.de}

\author{Nikolaus A. Adams}
\ead{nikolaus.adams@tum.de}

\address{Department of Mechanical Engineering, Chair of Aerodynamics and Fluid Mechanics, Technical University of Munich, Germany}

\begin{frontmatter}
\begin{abstract}
In this work, physics-informed neural networks are applied to incompressible two-phase flow problems. We investigate the forward problem, where 
the governing equations are solved from initial and boundary conditions, as well as the inverse problem, where continuous velocity and 
pressure fields are inferred from scattered-time data on the interface position. We employ a volume of fluid approach, i.e. the
auxiliary variable here is the volume fraction of the fluids within each phase. For the forward problem, we solve the 
two-phase Couette and Poiseuille flow. For the inverse problem, three classical test
cases for two-phase modeling are investigated: (i) drop in a shear flow, (ii) oscillating drop and (iii) rising bubble.
Data of the interface position over time is generated by numerical simulation.
An effective way to distribute spatial training points to fit the interface, i.e. the volume fraction field, and the residual points is proposed.
Furthermore, we show that appropriate weighting of losses associated with the residual of the partial differential equations is crucial for successful training.
The benefit of using adaptive activation functions is evaluated for both the forward and inverse problem.
\\
\textit{Keywords:} Physics-informed neural networks, Two-phase flows, Volume-of-fluid, Hidden fluid mechanics, Incompressible Navier-Stokes equations
\end{abstract}

\end{frontmatter}
\journal{MLWA}

\section{Introduction}
\label{section:1}

Since multiphase flows with fluid-fluid interfaces are omnipresent in nature and industrial applications, the accurate numerical simulation of such flows is subject of intense research.  
Despite the challenges posed by multiphase flows such as large density and material parameter ratios \citep{Bussmann2002, Scardovelli1999} and modeling of surface tension \citep{Popinet2018}, a wide range 
of interface tracking/capturing algorithms has been developed. Examples include the Level-set \citep{Sussman1994b}, Front-tracking \citep{Tryggvason2001} and Volume-of-fluid \citep{Hirt1981} method. 

Recent successes of machine learning in cognitive science \citep{Lake2015} and genomics \citep{Alipanahi2015}
have motivated researchers to apply such techniques to computational physics. Deep learning in particular has brought about major
breakthroughs in image and speech recognition \citep{Lecun2015}. The origin of deep learning methods for solving physical problems governed by 
partial differential equations (PDE) may be tracked back to \cite{Lagaris1998}. \cite{Raissi2019} introduced physics-informed neural networks (PINN): Deep neural networks that are constrained to respect the underlying pyhsical laws of the
observed data. The strength of this approach lies in its flexibility as it may be used for inference, i.e. data driven solution of PDE, but also for system identification, i.e. data driven discovery of PDE.
In the latter case, parameters of the governing equations may be completely or partially unkown, and are learned from the observed data. For inference, PINN provide an alternative approach to traditional mesh-based methods for solving PDE from initial and boundary conditions, referred to as the forward problem. However, they
also provide an arguably more attractive application, which is solving the inverse problem. In this case, potentially sparse and noisy data scattered across space and time of either auxiliary variables and/or some quantities of interest (QoI) are availabe. These data alongside the given physical laws are used to infer the entirety of the QoI within the whole spatio-temporal domain.       

The performance of PINN has been improved continuously. \cite{Lu2019} presented DeepXDE, which is a customizable python framework providing building blocks to construct individual problems regarding the spatio-temporal domain and boundary conditions. They proposed
a new method to distribute the training points for the residual of the PDE (residual points). Analogously to adaptive refinement in mesh-based solvers, residual points are added during training where the residua of the PDE is large, improving the efficiency of the training process. \cite{Dwivedi2019} proposed distributed PINN to evade the difficulty associated with training deep neural networks
due to the well known problem of unstable and vanishing gradients \citep{Pascanu2013}. Instead of using one potentially deep PINN for the whole spatio-temporal domain, the authors suggest to divide the domain into cells and locally install a PINN in each of these cells. To preserve global continuity and differentiability of the solution, it is necessary 
to minimize corresponding losses at the cell faces. Compared to one global PINN, the local PINN needs to perform a less complex task. Thus they require less hidden layers and are therefore easier to train.
\cite{Jagtap2019,Jagtap2020} suggest adaptive activation functions by introducing scaled coefficients into the activation functions, which are optimized alongside the network parameters resulting in better convergence and accuracy.

The possibility of solving the inverse problem naturally gives rise to novel approaches for the quantification of flow fields. Direct measurements of the velocity and pressure for an entire flow field is usually impractical.
The idea is to combine easier obtainable data of auxiliary variables with the knowledge of underlying physical laws to extract quantitative information of the entire flow field. 
One example is the inference of velocity and pressure fields from a passive scalar, e.g. smoke or contrast agents, that are used to experimentally visualize the flow field. 
By encoding the incompressible Navier-Stokes equations and an advection-diffusion equation for a passive scalar into a PINN, Raissi et al. have shown that hidden fluid mechanics \citep{Raissi2020, Raissi2019a} is a promising application. Another example was presented by \cite{Mao2020b}, who quantified compressible flow fields using data
of the density gradient, which experimentally may be generated using Schlieren photography. We extend this idea to two-phase flows using a Volume-of-fluid approach. Here, the neural network receives data on the interface position scattered over time,
motivated by experimental setups where the interface position is obtained using non-invasive methods, namely optical techniques combined with image processing \citep{Murai2001, Takamasa2003, Murai2006, Poletaev2020}, provided 
a direct observation of the interface is possible. In experiments that do not allow for direct observation, methods such as ultrasonic detection \citep{Murai2010} may be used. Best-practice precedents for training point distribution and choice of hyper-parameters for the neural network and training process are shown.
The predicted results are validated with analytical and CFD solutions. The CFD simulations are performed with ALPACA \citep{Winter2019, Kaiser2019, Kaiser2019a}, a sharp interface capturing (Level-set) two-phase solver for compressible flows.

The paper is structured as follows: Section \ref{section:2} provides a description of physics-informed neural networks and the governing equations including the Volume-of-fluid method. 
In section \ref{section:3} the results are presented. For the forward problem, a 1D two-phase Couette flow and a quasi 1D two-phase Poiseuille flow is investigated. Subsequently, the inverse problem is studied on various 2D test cases including a drop in a shear flow, an oscillating drop and a buoyancy driven rising bubble. Section \ref{section:4} concludes this work with final remarks.


\section{Methodology}
\label{section:2}

\subsection{Governing equations}
The governing equations for an incompressible, Newtonian fluid under the effect of surface tension and gravitational forces in non-dimensional ($^*$) form are given by
\begin{align}
    \nabla^* \cdot \textbf{u}^* &= 0 \label{eq:continuity}, \\
    \rho^*\left(\frac{\partial\textbf{u}^*}{\partial t^*} + (\textbf{u}^*\cdot\nabla^*)\textbf{u}^*\right) &= \nonumber \\
    -\nabla^* p^* &+ \nabla^*\cdot\frac{1}{Re} (\nabla^* \textbf{u}^* + \nabla^* \textbf{u}^{*T}) + \frac{1}{We}\kappa \nabla^* \alpha+ \rho^*\frac{1}{\textbf{Fr}^2}, \label{eq:momentum}
\end{align}
where $\textbf{u}^*=[u^*,v^*]^T$ and $p^*$ and $\rho^*$ denote the velocity, pressure and densitiy, respectively. The Reynolds number $Re$,
Weber number $We$ and Froude number $\mathbf{Fr}$ are defined as
\begin{align}
    Re&=\frac{\rho_r u_r L_r}{\mu},\\
    We&=\frac{\rho_r u_r^2 L_r}{\sigma},\\
    \textbf{Fr}&=\frac{u_r}{\sqrt{\textbf{g} L_r}},
\end{align}
with $\rho_r$, $u_r$ and $L_r$ being reference quantities for the density, velocity and length scale.
The dynamic viscosity, surface tension coefficient and gravitational acceleration
are indicated by $\mu$, $\sigma$ and $\mathbf{g}$, respectively. The relations between dimensional and non-dimensional quantities are as follows:
$\mathbf{u}=\mathbf{u}^*u_r$, $\rho=\rho^*\rho_r$, $\mathbf{x}=\mathbf{x}^*L_r$ and $p=p^*\rho_ru_r^2$. 

In the Volume-of-fluid \citep{Hirt1981} method, $\alpha$ represents the volume fraction of the respective fluid in each computational cell. Within the proposed deep learning framework the background computational mesh is used. In this case a phase characteristic variable $\phi$ \citep{Scardovelli1999}, 
which can only adopt the values 0 and 1, identifies the phase assignment of any point of the domain. The relation between both scalar fields is $\alpha=\sfrac{\int \phi \mathrm{dV}}{\Delta V}$, where $\Delta V$ is the volume of a computational cell. The local averaged density $\rho$ and viscosity $\mu$ 
are evaluated as
\begin{align}
    \rho &= \rho_2 + (\rho_1 - \rho_2)\alpha, \\
    \mu &= \mu_2 + (\mu_1 - \mu_2)\alpha. 
\end{align} 
Note that fluid $1$ and fluid $2$ correspond to a volume fraction of $1$ and $0$, respectively. The interface region is indicated by $0<\alpha<1$. The volume fraction is advected by the local fluid velocity.
\begin{equation}
    \frac{\partial \alpha}{\partial t^*} + \textbf{u}^*\cdot\nabla^* \alpha = 0. \label{eq:advection}
\end{equation}
The curvature is given by
\begin{align}
    \kappa = - \nabla^* \cdot \frac{\nabla^* \alpha}{|\nabla^* \alpha|}.     
\end{align}

\subsection{Physics-informed neural networks}
Let $\mathbf{U}(\mathbf{x}^*,t^*)=[\mathbf{u}^*, p^*, \alpha]^T$ be the solution vector of the set of equations \eqref{eq:continuity}, \eqref{eq:momentum} and \eqref{eq:advection}.
The solution is approximated by a deep neural network $\hat{\mathbf{U}}(\mathbf{x}^*, t^*; \mathbf{w})$, with $\mathbf{w}$ being the network parameters, namely its weights and biases, that are 
tuned during the training process. For a detailed description of how deep neural networks function regarding feed forward and backpropagation, the reader is referred to \citep{Goodfellow2016}.
Using automatic differentation \citep{GunesBaydin2018, abadi2016tensorflow}, $\hat{\mathbf{U}}$ may be derived with respect to its inputs. This way, the residual network
\begin{equation*}
    \hat{\mathbf{f}}(\mathbf{x}^*,t^*;\mathbf{w}) = \left[
    \begin{array}{c}
        \nabla^* \cdot \hat{\mathbf{u}}^* \\
        \rho^*\left(\frac{\partial\hat{\textbf{u}}^*}{\partial t^*} + (\hat{\textbf{u}}^*\cdot\nabla^*)\hat{\textbf{u}}^*\right) +\nabla^* \hat{p}^* - \nabla^*\cdot\frac{1}{Re} (\nabla^* \hat{\textbf{u}}^* + \nabla^* \hat{\textbf{u}}^{*T}) - \frac{1}{We}\kappa \nabla^* \hat{\alpha} - \rho^*\frac{1}{\textbf{Fr}^2} \\
        \frac{\partial \hat{\alpha}}{\partial t^*} + \hat{\textbf{u}}^*\cdot\nabla^* \hat{\alpha}
    \end{array}\right]
\end{equation*}
that maps the spatio-temporal coordinates to the residuals of the governing equations described above, is defined. Note that $\hat{\mathbf{U}}$ and $\hat{\mathbf{f}}$ share weights and biases.
During training, $\hat{\mathbf{U}}$ receives the observed data, i.e. initial and boundary conditions in the forward problem and scattered measurements of 
auxiliary variables or partial quantities of interest in the inverse problem. Simultaneously, the residual network $\hat{\mathbf{f}}$ acts as a regularization agent for possible solutions that fit the observed data to also fulfill 
the governing equations. We minimize the mean squared error between prediction and observed data $MSE_u$ and the mean squared error of the residual network output $MSE_f$.
\begin{align}
    MSE_u &= \frac{1}{N_u} \sum_{i=1}^{N_u} |\mathbf{U}^i - \hat{\mathbf{U}}(\mathbf{x}_u^i, t_u^i)|^2 \label{eq:MSE_u}\\
    MSE_f &= \sum_j w_{f,j} \frac{1}{N_f} \sum_{i=1}^{N_f} |\hat{f}_j(\mathbf{x}_f^i, t_f^i)|^2, \quad \mathrm{with} \quad j=\{m,u,v,\alpha\} \label{eq:MSE_f}
\end{align}
Here, $\{\mathbf{U}^i, \mathbf{x}_u^i, t_u^i\}|_{i=1}^{N_u}$ and $\{\mathbf{x}_f^i, t_f^i\}|_{i=1}^{N_f}$ denote the set of training points for the observed data and the set of residual points, respectively.
The corresponding number of samples is indicated by $N_u$ and $N_f$. Two important notes shall be made. First, $MSE_u$ is not yet explicitly defined as it is in practice a sum of multiple losses that are associated with observed data.
These might e.g. be boundary/initial conditions and/or scattered measurements across the spatio-temporal domain. This implies that the composition of $MSE_u$ is case dependent and needs to be specified for each case individually.
Second, $MSE_f$ is a weighted sum of the mean squared error of the continuity, momentum and volume fraction advection equation residuals 
\begin{equation}
    \label{eq:weighting}
    MSE_f = w_{f,m}MSE_{f,m} + w_{f,u}MSE_{f,u} + w_{f,v}MSE_{f,v} + w_{f,\alpha}MSE_{f,\alpha}.
\end{equation} 
The indices $m,u,v$ and $\alpha$ indicate continuity, $x$- and $y$-momentum and advection equation, respectively. The loss weights are crucial for the training process when inferring flow fields with surface tension forces as will be shown later.
They are case dependend and will thus be specified in the results for each case individually. Note that the components $MSE_{f,i}$ all are computed at the same residual points $\{\mathbf{x}_f^i, t_f^i\}|_{i=1}^{N_f}$.

Adaptive activation functions \citep{Jagtap2019,Jagtap2020} introduce a scaled coefficient $n\cdot a$ into the activation function argument. Here, $n$ is the scale factor and $a$ 
is the adaptive activation coefficient, which is optimized along with the neural network parameters. The scale factor is used to adjust the sensitivity of the optimization process with respect to $a$. 
The initial values of $a$ and $n$ are case dependend and will thus be specified for each case individually in the results.

\section{Results}
\label{section:3}
To quantify the error between neural network predictions and analytical or CFD solutions (both will be referred to as exact solution), the mean absolute error $MAE$ and the relative $L_1$ and $L_2$ 
error norms will be used. For an arbitrary quantity $q$, these errors are computed as follows:
\begin{subequations}
    \label{erorrnorms}
\begin{align}
    MAE &= \frac{1}{N} \sum_i^N |q_{i,nn} - q_{i,exact}|, \\
    L_{1} &= \frac{\sum_i^N |q_{i,nn} - q_{i,exact}|}{\sum_i^N |q_{i,exact}|}, \\    
    L_{2} &= \frac{\sqrt{\sum_i^N (q_{i,nn} - q_{i,exact})^2}}{\sqrt{\sum_i^N q_{i,exact}^2}}.    
\end{align}
\end{subequations}

For the following test cases, weights and biases are initialized by glorot normal \citep{Glorot2010} and zero initialization, respectively. The activation functions
for the hidden layers are hyperbolic tangents. The output layer activation functions are linear for the velocities. For the 
pressure and the volume fraction, an exponential and logistic function is used, respectively. As for the training procedure, the network parameters are optimized using the Adam algorithm \citep{Kingma2015}.
The training data are shuffled at the start of each epoch and subsequently split into multiple batches. The training and neural network hyper-parameters will be specified for each problem, individually. 
The implementation is done with the library TensorFlow \citep{abadi2016tensorflow}. \textcolor{red}{We provide the code to train and visualize the rising bubble case at} \url{https://github.com/aaronbuhendwa/twophasePINN}.

\subsection{Forward Problems}
The forward problem is characterized by solving the partial differential equation from initial and boundary conditions.
Thus, the total loss function $MSE_{total}$ here consists of the losses associated with enforcing the initial and boundary conditions $MSE_{IC}$ and $MSE_{BC}$ plus the loss
of the residual of the governing equations $MSE_f$.
\begin{equation}
    MSE_{total} = MSE_{IC} + MSE_{BC} + MSE_f \label{MSE_total_forward}
\end{equation}
\subsubsection{Couette Flow}
\begin{figure}[!t]
    \centering
    \input{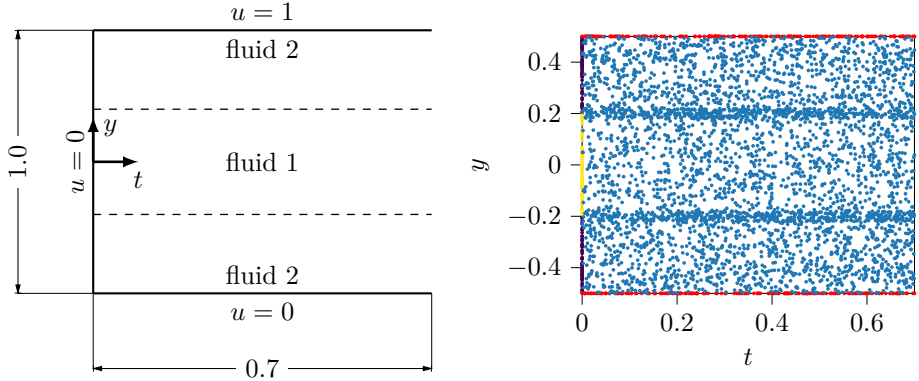}
    \caption{(Left) Schematic of the initial and boundary conditions for the two-phase Couette flow. (Right) Corresponding point distribution. 
    Yellow and violet marker colors represent the points to enforce the initial condition and indicate a volume fraction of 1 and 0, respectively. Blue markers represent the residual points and red markers indicate points for the boundary conditions.}
    \label{fig:couette_setup_points}
\end{figure}

The Couette flow describes a viscous fluid that is shear driven by one or multiple moving no slip boundaries. In the present two-phase configuration, the two fluids are bound by the spatio-temporal domain $y \in [-0.5,0.5]$ and $t \in [0.0,0.7]$ 
with interfaces at $y=-0.2$ and $y=0.2$, see Figure \ref{fig:couette_setup_points} on the left. The analytical steady state solution for this setup is a piecewise linear velocity profile with
kinks at the interfaces, as derived using boundary conditions and the balance of shear stresses $\mu_1 \frac{\partial u_1}{\partial y} = \mu_2 \frac{\partial u_2}{\partial y}$ 
at both interfaces. Note that the indices 1 and 2 denote properties of fluid 1 and 2, respectively. The boundary conditions are $u(0.5,t) = 1.0$ and $u(-0.5,t) = 0.0$ and the initial condition is $u(y,0)=0.0$ and
\begin{equation*}
    \alpha(y,0) = 
    \begin{cases} 
        0       & \text{if } |y| \geq 0.2 \\
        1       & \text{otherwise}
    \end{cases}
\end{equation*}
All reference quantities for non-dimensionalization are 1.0.

The loss functions for initial and boundary conditions are 
\begin{align*}
    MSE_{IC} &= \frac{1}{N_{IC}} \sum_{i=1}^{N_{IC}} |\alpha^i_{IC} - \hat{\alpha}(y^i_{IC}, t_{IC}^i)|^2 + |u^i_{IC} - \hat{u}(y^i_{IC}, t_{IC}^i)|^2 \\
    MSE_{BC} &= \frac{1}{N_{BC}} \sum_{i=1}^{N_{BC}} |u^i_{BC} - \hat{u}(y^i_{BC}, t^i_{BC})|^2
\end{align*}
where $\{\alpha^i_{IC}, u^i_{IC}, y_{IC}^i, t_{IC}^i\}|_{i=1}^{N_{IC}}$ represents the data for the initial condition and $\{u^i_{BC}, y_{BC}^i\}|_{i=1}^{N_{BC}}$ corresponds to the data for the boundary condition. The loss weights are $w_{f,i}=1.0$ with $i = \{m,u,v,\alpha\}$.
Note that $MSE_f$ is computed as described by equations \eqref{eq:MSE_f} and \eqref{eq:weighting}.

The training points are drawn from a random uniform distribution within the respective bounds. Figure \ref{fig:couette_setup_points} shows the point distribution for this case. Notice the refinement region for the residual points
close to the interface. This region spans 0.02 in positive and negative $y$-direction from each interface. 200 points are distributed for the initial condition while 400 points are used in total for the boundary conditions, i.e. 200 points for each boundary.
This results in $N_{IC}=200$ and $N_{BC}=400$. The spatio-temporal domain is filled with 3000 residual points excluding the interface region. The latter is refined with 1000 points, i.e. 500 for each interface. 
The set of points for the initial and boundary conditions are added to the set of residual points, ensuring that the residual of the PDE are also minimized at $t=0$ and the boundaries. This results in $N_f=4600$.

\begin{table}[!b]
    \centering
    \setlength\arrayrulewidth{1pt}
    \begin{tabular}{lccccc}
        \hline
        epochs& 500 & 2000 & 2000 & 2000 & 5000 \\
        learning rate & \num{5e-4} & \num{1e-4} & \num{5e-5} & \num{1e-5} & \num{5e-6}\\
        batches & 10 & 10 & 10 & 10 & 10\\
        \hline
        \end{tabular}
    \caption{Training hyper-parameters for the two-phase Couette flow.}
    \label{table:couette_hyperparameters}  
\end{table}

The neural network used for this case maps $y,t \rightarrow u$ and consists of 10 hidden layers with 150 nodes each. The terms in equations \eqref{eq:continuity}, \eqref{eq:momentum} and \eqref{eq:advection}
that may be neglected for this specific case are not implemented for computational time reasons. In particular, these are terms with $v$, $\frac{\partial}{\partial x}$ and gravitational as well as surface tension forces.
From experience, it has proven to be useful to successively reduce the learning rate during the training procedure. Table \ref{table:couette_hyperparameters} lists the training hyper-parameters
for this case.

\begin{figure}[!t]
    \centering
    \input{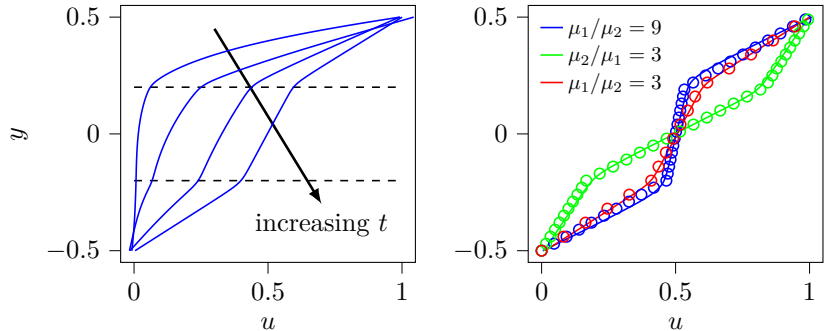}
    \caption{(Left) Predicted time evolution for the two-phase Coeutte flow with $\mu_1/\mu_2=3$ for time snapshots $t=0.02, 0.05, 0.12, 0.05$. (Right) Comparison of the steady state solution of the two-phase Couette flow for different viscosity ratios. Predicitons and analytical solutions are indicated by lines and markers, respectively.}
    \label{fig:couette_mu_variation}
\end{figure}

\begin{table}[!t]
    \centering
    \setlength\arrayrulewidth{1pt}
    \begin{tabular}{ccccc}
        \hline
        $\mu_1 / \mu_2$ & 9 & 6 & 3 & 1/3 \\
        \hline
        rel. $L_2$ error & \num{3.56e-2} & \num{2.27e-2} & \num{1.47e-2} & \num{1.62e-2} \\
        rel. $L_1$ error & \num{2.98e-2} & \num{1.96e-2} & \num{1.39e-2} & \num{1.69e-2} \\
        MSE & \num{3.87e-4} & \num{1.59e-4} & \num{6.81e-5} & \num{9.78e-05} \\
        \hline
    \end{tabular}
    \caption{Steady state two-phase Couette flow errors to the analytical solution for different viscosity ratios.}  
    \label{table:couette_variation_errors}
\end{table}

In Figure \ref{fig:couette_mu_variation} the predicted time evolution for a fixed viscosity ratio is shown on the left.
Furthermore, the predicted and exact steady state solution for various viscosity ratios are compared on the right. To evaluate the error between prediction and analytical solution,
a linearly spaced grid of 100 spatial points in $y$-direction is used. The corresponding errors are presented in Table \ref{table:couette_variation_errors},
with relative errors ranging between $1.4\%$ to $3.6\%$, increasing with the viscosity ratio. Note that these results are for fixed activation functions. 

\begin{figure}[!t]
    \centering
    \input{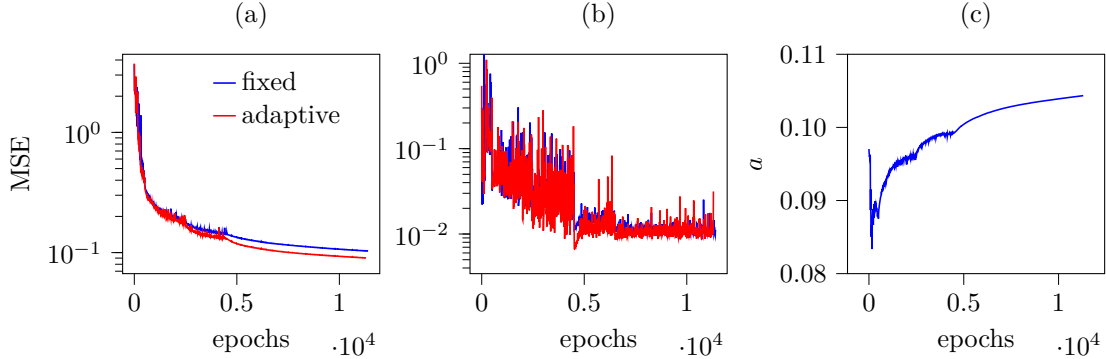}
    \caption{Comparison of the training history of the two-phase Couette flow with $\mu_1/\mu_2=3$ for fixed and adaptive activation functions. (a) Losses associated with the initial and boundary conditions. (b) Losses associated with the PDE residual. (c) Adaptive activation coefficient.}
    \label{figure:couette_loss_history}
\end{figure}

\begin{table}[!b]
    \centering
    \setlength\arrayrulewidth{1pt}
    \begin{tabular}{cccr}
        \hline
        & fixed & adaptive \\
        \hline
        rel. $L_2$ error & \num{1.62e-2} & \num{1.26e-2}\\
        rel. $L_1$ error & \num{1.69e-2} & \num{1.32e-2}\\
        MSE & \num{9.78e-05} & \num{5.87e-05} \\
        \hline
    \end{tabular}
    \caption{Steady state two-phase Couette flow errors to the analytical solution for $\mu_2/\mu_1 = 3$ for fixed and adaptive activation functions.}
    \label{table:couette_errors_ad_act}  
\end{table}

To study the influence of adaptive activation functions, a single adaptive activation coefficient $a$ is used for all nodes of the hidden layers. An initial value of $0.1$ with a scale factor of $10$ is chosen. These values have been found
to be decent regarding convergence rate improvements and the sensitivity of $a$ \citep{Jagtap2019}. In Figure \ref{figure:couette_loss_history} the training history for with and without adaptive activation functions for a fixed viscosity ratio is depicted. 
An initial drop of $a$ is observed. After the first learning rate reduction at $\mathrm{epoch}=500$ the adapative activation coefficient starts to increase, accompanied with a better convergence rate of the losses. 
Note the distinct jumps of both the losses and $a$ at certain epochs. This is due to the successive reduction of the learning rate. In Table \ref{table:couette_errors_ad_act} the steady state errors with and without adaptive activation functions are shown.
A reduction of the steady state relative errors of about $20\%$ is achieved. At the same time, the computational cost per epoch is increased by about $28\%$. 
\subsubsection{Poiseuille flow}
\begin{figure}[!b]
    \centering
    \begin{tikzpicture}
        \coordinate (A) at (0,4);
        \coordinate (B) at (0,0);
        \draw[line width=1pt, fill=black!20!white] (A) rectangle (4,4.1);
        \draw[line width=1pt, fill=black!20!white] (B) rectangle (4,-0.1);
        \node[below] at (0.7,3.9) {fluid 1};
        \node[above] at (0.7,0.1) {fluid 2};
        \draw[dashed] (0,2) -- (4,2);
        \draw[line width=0.5] (0,0) -- (0,4);
        \draw[line width=0.5] (4,0) -- (4,4);
        \draw[->, line width=1] (0.3,2.5) -- (1.7,2.5) node[midway, above] {$\frac{\partial p}{\partial x} < 0 $};
        \draw[->, line width=1] (2,2) -- (2.6,2) node[at end, below] {$x$};
        \draw[->, line width=1] (2,2) -- (2,2.6) node[at end, right] {$y$}; 
        \dimline[extension start length=1cm, extension end length=1cm,extension style={black}, label style={rotate=-90}] {(-1, 0)}{(-1, 4)}{1.0};
        \dimline[extension start length=-1cm, extension end length=-1cm, extension style={black}] {(0,-1)}{(4,-1)}{0.1};
        \dimline[extension start length=0cm, extension end length=0cm, extension style={black}] {(3,0)}{(3,2)}{$b$};
        \dimline[extension start length=0cm, extension end length=0cm, extension style={black}] {(3,2)}{(3,4)}{$b$};
        \node[rotate=90, above] at (-0.1,2) {periodic, $p=3.4$};
        \node[rotate=90, below] at (4.1,2) {periodic, $p=1.0$};
        \node[above] at (2,4.1) {$u=0$,$p=f(x)$}; 
        \node[below] at (2,-0.1) {$v=0$,$p=f(x)$}; 
    \end{tikzpicture}
    \caption{Schematic of initial and boundary conditions of the two-phase Poiseuille flow.}
    \label{fig:poiseuille_setup}
\end{figure}
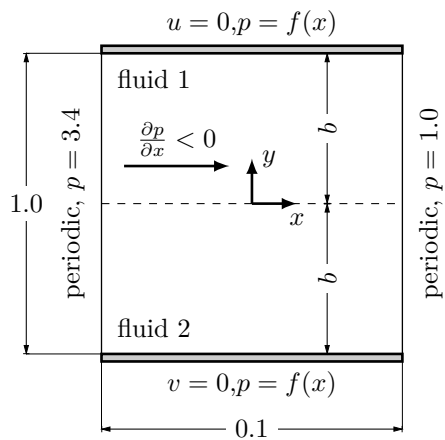
The poiseuille flow describes a viscous channel flow driven by a constant pressure gradient. The quasi 1D two-phase configuration studied here is depicted in Figure \ref{fig:poiseuille_setup}. The two fluids are 
bound by the spatio-temporal domain $x\in [-0.1,0.1]$, $y\in [-0.5,0.5]$ and $t\in [0.0,0.5]$ with an interface at $y=0.0$. At the north and south boundary, the no slip condition $u=0$ and a linear pressure profile from 
west value $p_W=3.4$ to east value $p_E=1.0$ is prescribed. This results in a pressure gradient of $\frac{\partial p}{\partial x}=-12$. The east and west boundaries are periodic for the velocity and fixed value for the pressure.
The initial conditions are $u(x,y,0) = 0$, $p(x,y,0)=p_W + \frac{\partial p}{\partial x}x$ and 
\begin{equation*}
    \alpha(x,y,0) = 
    \begin{cases} 
        1       & \text{if } y \geq 0.0 \\
        0       & \text{otherwise}
    \end{cases}
\end{equation*}
The analytical steady state solution for this setup consists of two quadratic functions for each phase with a kink at the interface. It is given by \citep{Rezavand2018a}
\begin{equation}
    u_i = \frac{1}{\rho}\frac{\partial p}{\partial x} \frac{b^2}{2\mu_i}\left[ \left(\frac{2\mu_i}{\mu_1+\mu_2}\right) + \left(\frac{\mu_1 - \mu_2}{\mu_1 + \mu_2}\right)\left(\frac{y}{b}\right) - \left(\frac{y}{b}\right)^2\right],
\end{equation}
where $u_i$ with $i\in\{1,2\}$ denotes the steady state velocity profile within the respective fluid and $b=0.5$ is half the domain height. 
All reference quantities for non-dimensionalization are 1.0.

The loss functions for initial and boundary conditions are given by
\begin{alignat*}{2}
    MSE_{IC} &= \frac{1}{N_{IC}} \sum_{i=1}^{N_{IC}} |\alpha^i - \hat{\alpha}(\mathbf{x}_{IC}^i, t_{IC}^i)|^2 + |u^i - \hat{u}(\mathbf{x}_{IC}^i, t_{IC}^i)|^2 + |p^i - \hat{p}(\mathbf{x}_{IC}^i, t_{IC}^i)|^2, \\
    MSE_{NS} &=  \frac{1}{N_{NS}} \sum_{i=1}^{N_{NS}} |u^i_{NS} - \hat{u}(\mathbf{x}^i_{NS},t^i_{EW})|^2 + |p^i_{NS} - \hat{p}(\mathbf{x}^i_{NS},,t^i_{NS})|^2, \\
    MSE_{EW} &=  \frac{1}{N_{EW}} \sum_{i=1}^{N_{EW}} |\hat{u}(x^i_W,y^i_{EW}, t^i_{EW}) - \hat{u}(x^i_E,y^i_{EW}, t^i_{EW})|^2 + |p^i_{EW} - \hat{p}(x^i_{EW},y^i_{EW}, t^i_{EW})|^2, \\
    MSE_{BC} &= MSE_{EW} + MSE_{NS},
\end{alignat*}
where $\{\alpha^i_{IC}, u^i_{IC}, \mathbf{x}_{IC}^i, t_{IC}^i\}|_{i=1}^{N_{IC}}$ denotes the data for the initial condition, $\{p^i_{NS}, u^i_{NS}, \mathbf{x}_{NS}^i, t_{NS}^i\}|_{i=1}^{N_{NS}}$ is the training data for the north and south
boundary and $\{p^i_{EW}, x^i_E, x^i_W, y_{EW}^i, t_{EW}^i\}|_{i=1}^{N_{EW}}$ represents the data for the east and west boundary.
The loss weights are $w_{f,i}=1.0$ with $i = \{m,u,v,\alpha\}$. Note that $MSE_f$ is computed as described by equations \eqref{eq:MSE_f} and \eqref{eq:weighting}.

\begin{figure}[!b]
    \centering
    \input{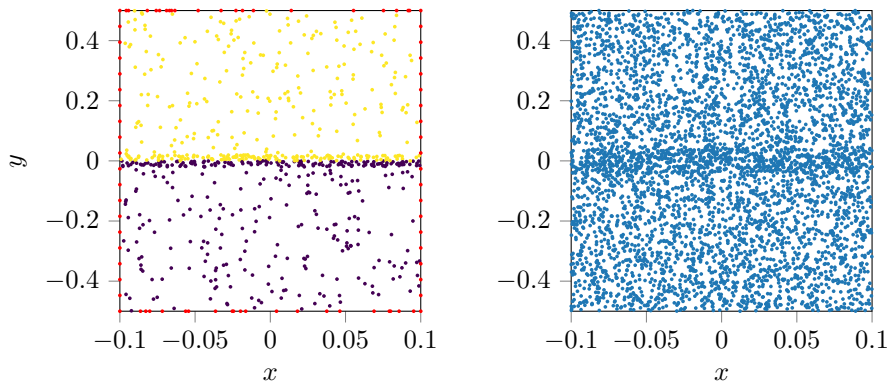}
    \caption{Point distribution for the two-phase Poiseuille flow at $t=0$. (Left) Points for the initial and boundary conditions. Yellow and violet marker colors indicate a volume fraction of 1 and 0, respectively. Red markers indicate points for the boundary conditions.
    (Right) Residual points.}
    \label{fig:poieuille_points}
\end{figure}

\begin{figure}[!b]
    \centering
    \begin{tikzpicture}
    \node at (-3.5,1.7) {\includegraphics[scale=0.37]{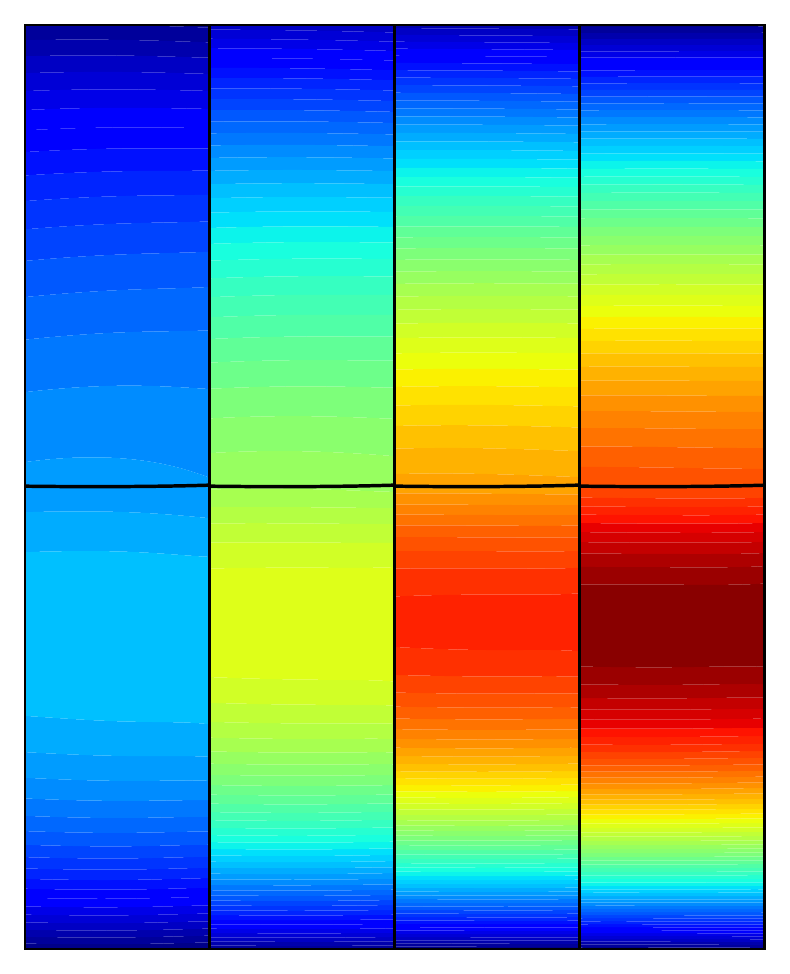}};
    \input{images/poiseuille_mu_variation.tex}
    \end{tikzpicture}
    \caption{(Left) Qualitative time evolution of the two-phase Poiseuille with $\mu_1/\mu_2=4$ for time snapshots $t=0.02, 0.05, 0.1, 0.5$ predicted by the neural network. Blue and red color color indicate minimum and maximum value within the 
    whole spatio-temporal domain. (Right) Comparison of the steady state solution of the two-phase Poiseuille flow for different viscosity ratios at $y=0.0$ with $\bar{u}$ being the corresponding mean velocity. Predicitons and analytical solutions are indicated by lines and markers, respectively.}
    \label{fig:poiseuille_variation}
\end{figure}

The distribution of the training points for $t=0$ is shown in Figure \ref{fig:poieuille_points}. For the initial condition, the interface is refined in a range of 0.02 in both positive and negative $y$-direction from the interface with 300 points. For the rest of the domain,
400 points are used, resulting in $N_{IC} = 700$. We draw 20 time snapshots randomly from a uniform-distribution within the respective bound to distribute the spatial points for each boundary and the residual points.
Here, it is ensured that $t=0.0$ and $t=0.5$ are included in the set of time snapshots and that
the time snapshots for the east and west boundary are equal, since this is necessary for the periodic condition. At each time snapshot, the interface is refined with 800 spatial residual points. The refinement region for the residual points spans 0.04 in positive and negative $y$-direction from the interface.
The rest of the domain is filled with 4000 spatial residual points. As for the boundary conditions, 20 spatial points are random uniformly drawn at the north and south boundary and linearly spaced at the east and west boundary within 
the respective bounds, resulting in $N_{NS} = 800$ and $N_{EW} = 400$, respectively. Note that for periodic boundaries, one sample corresponds to a pair of points.
As described in the previous case, the set of points for the initial and boundary conditions are added to the set of residual points. This results in a total of $N_f=98300$ points. 

\begin{table}[!t]
    \centering
    \setlength\arrayrulewidth{1pt}
    \begin{tabular}{lccccc}
        \hline
        epochs& 3000 & 3000 & 3000 & 3000 & 10000 \\
        learning rate & \num{5e-4} & \num{1e-4} & \num{5e-5} & \num{1e-5} & \num{5e-6}\\
        batches & 10 & 10 & 10 & 10 & 10\\
        \hline
    \end{tabular}
    \caption{Training hyper-parameters for the two-phase Poiseuille flow.}
    \label{table:poiseuille_hyperparameters}  
\end{table}

For this case, a neural network mapping $x,y,t \rightarrow u, p, \alpha$ consisting of 10 hidden layers with 100 nodes each is used. We again neglect terms in equations \eqref{eq:continuity}, \eqref{eq:momentum} and \eqref{eq:advection} that 
are not needed, i.e. terms with $v$ and gravitational/capillary forces. Table \ref{table:poiseuille_hyperparameters} lists the training hyper-parameters. 
\begin{table}[!t]
    \centering
    \setlength\arrayrulewidth{1pt}
    \begin{tabular}{ccccc}
        \hline
        $\mu_1 / \mu_2$ & 2 & 4 & 6 & 8 \\
        \hline
        relative $L_2$ error & \num{5.67e-3} & \num{8.61e-3} & \num{1.32e-2} & \num{4.42e-2} \\
        relative $L_1$ error & \num{5.95e-3} & \num{8.04e-3} & \num{1.26e-2} & \num{3.39e-2} \\
        MSE & \num{1.81e-5} & \num{1.86e-5} & \num{2.79e-5} & \num{2.36e-4} \\
        \hline
    \end{tabular}
    \caption{Steady state two-phase Poiseuille flow errors of the prediction to the analytical solution for different viscosity ratios.}  
    \label{table:poiseuille_variation_errors}
\end{table}
\begin{figure}[!b]
    \centering
    \input{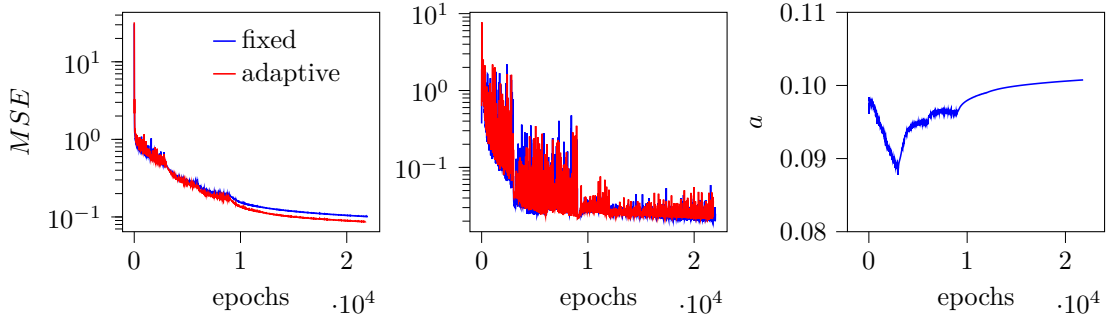}
    \caption{Comparison of the training history of the two-phase Poiseuille flow with $\mu_1/\mu_2=4$ for fixed and adaptive activation functions. (Left) Losses associated with the initial and boundary conditions. (Middle) Losses associated with the PDE residual. (Right) Adaptive activation coefficient.}
    \label{fig:poiseuille_loss_history}
\end{figure}
In Figure \ref{fig:poiseuille_variation} the predicted time evolution for a fixed viscosity ratio is shown on the left. On the right side, the predicted and exact steady state solutions for various viscosity ratios are compared. 
Table \ref{table:poiseuille_variation_errors} lists the corresponding errors. The relative errors are ranging from $0.55\%$ to $4.4\%$ increasing with the viscosity ratio $\mu_1/\mu_2$. These errors have been evaluated on a linearly spaced grid of 100 points in $y$-direction at $x=0$. 

A single adaptive activation coefficient $a$ is introduced for each node of the hidden layers to study its influence on the training process. The initial value is $0.1$ and a scale factor of $10$ is used. 
Figure \ref{fig:poiseuille_loss_history} compares the loss history with and without adaptive activation functions for a fixed viscosity ratio. The observed pattern is quite similar to the previous case with an initial drop of $a$.
Subsequently, after the first learning rate reduction at $\mathrm{epoch}=3000$, an increase of both $a$ and the convergence rate of the losses can be observed. The usage of adaptive activation functions reduced the steady state relative errors 
to the analytical solution by about $51\%$, while the computational time per epoch increases by $48\%$.

\begin{table}[!t]
    \centering
    \setlength\arrayrulewidth{1pt}
    \begin{tabular}{ccc}
        \hline
        & fixed & adaptive \\
        \hline
        rel. $L_2$ error & \num{8.61e-3} & \num{4.10e-3}\\
        rel. $L_1$ error & \num{8.04e-3} & \num{3.47e-3}\\
        MSE & \num{1.86e-5} & \num{4.23e-6} \\
        \hline
    \end{tabular}
    \caption{Steady state two-phase Poiseuille flow errors to the analytical solution for $\mu_2/\mu_1 = 4$ for fixed and adaptive activation functions.}  
    \label{table:poiseuille_ad_act}
\end{table}

\subsection{Inverse Problems}
While the direct measurement of velocity and pressure of entire flow fields is impractical, the measurement of auxiliary variables like e.g. a passive scalar (smoke or dye) to visualize the flow
is comparatively easy. Combining these data of experimentally obtainable auxiliary variables with the knowledge of the underlying physical laws to extract quantitative information 
of the entire flow field is the motivation for the inverse problem.
Here, the auxiliary variable is the interface position, i.e. the volume fraction field, which in an experimental setup is obtained using a combination of optical techniques and image processing \citep{Murai2001,Murai2006,Takamasa2003, Poletaev2020}.
Three 2D test cases are investigated, namely a drop in a shear flow, an oscillating drop and a rising bubble. These are simulated with ALPACA \citep{Winter2019,Kaiser2019,Kaiser2019a}, which is a finite-volume based solver for compressible two-phase flows that captures the interface using the Level-set method \citep{Sussman1994b}. 
From the CFD solution, the information of the interface position over time is retrieved. The neural network predictions are then compared to the CFD solution.

\begin{figure}[!t]
    \centering
    \begin{tikzpicture}
        
    \draw[line width=2pt] (0,0) arc (0:15:20) node[pos=0.0] (A) {} node[pos=0.1] (A1) {} node[pos=0.2] (A2) {} node[shape=coordinate,pos=0.4] (A3) {} node[shape=coordinate,pos=0.3] (A4) {};
    \draw[line width=1pt, dashed] ({-(20-20*cos(3))},{-sin(3)*20}) arc (-3:18:20) node[pos=0.65, sloped, below] {interface};
    \draw[color=red] (0.8,0) arc (0:15:20.8) node[pos=0.0] (B) {} node[shape=coordinate,pos=1.0] (B1) {} node[shape=coordinate,pos=0.6] (B2) {};
    \draw[color=blue] (2,0) arc (0:15:22) node[pos=0.1] (C) {} node[shape=coordinate,pos=0.805] (C1) {};
    \draw[color=black!50!green] (4,0) arc (0:15:24) node[pos=0.2] (D) {} node[shape=coordinate,pos=0.8] (D1) {};

    \dimline[extension start length=0cm, extension end length=0cm, line style={color=red}, label style={above}] {(A)}{(B)}{$\delta_1$};
    \dimline[extension start length=0cm, extension end length=0cm, line style={color=blue}, label style={pos=0.75,above}] {(A1)}{(C)}{$\delta_2$};
    \dimline[extension start length=0cm, extension end length=0cm, line style={color=black!50!green}, label style={pos=0.75, above}] {(A2)}{(D)}{$\delta_3$};
    \filldraw[fill=green!20!white, draw=green!50!black] (D1) arc (12:15:24) -- (B1) arc (15:12:20) -- (D1);
    \filldraw[fill=blue!20!white, draw=blue] (C1) arc (12.075:9:22) -- (B2) arc (9:12.125:20.8) -- (C1);
    \filldraw[fill=red!20!white, draw=red] (B2) arc (9:6:20.8) -- (A3) arc (6:9:20) -- (B2);

    \draw[fill=green!20!white, draw=green!50!black] (5,5) rectangle (5.5,5.5) node[yshift=-0.25cm, align=left, right, font=\linespread{1}\selectfont] {nearfield refinement \\ region for residual points};
    \draw[fill=blue!20!white, draw=blue] (5,3) rectangle (5.5,3.5) node[yshift=-0.25cm, align=left, right, font=\linespread{1}\selectfont] {interface refinement \\ region for points to fit $\alpha$};
    \draw[fill=red!20!white, draw=red] (5,1) rectangle (5.5,1.5) node[yshift=-0.25cm, align=left, right, font=\linespread{1}\selectfont] {interface refinement \\ region for residual points};

    \node[below] at (1,-0.1) {fluid 1};
    \node[below] at (-1,-0.1) {fluid 2};

    \draw[draw=none] (-1,0) arc (0:15:19) node[pos=0.3,shape=coordinate] (E) {};
    \draw[->, line width=1pt] (A4) -- (E) node[at end, below] {$\textbf{n}$};

\end{tikzpicture}
    \caption{Schematic of the refinement regions for training point distribution.}
    \label{fig:interface_refinement}
\end{figure}
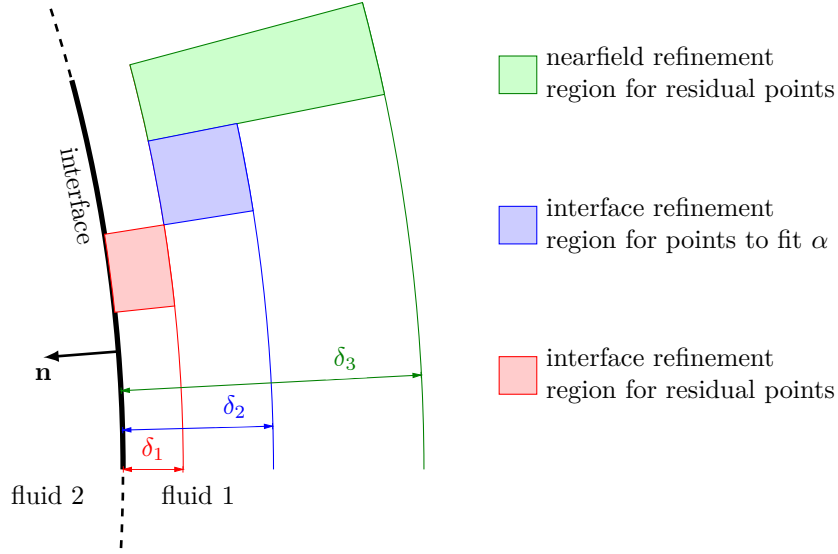

The interface refinement strategy with regard to the training point distribution is more sophisticated compared to what has been shown in the previous subsection for the 1D problems, since surface tension effects have now to be considered. 
In Figure \ref{fig:interface_refinement}, a representative interface segment with all refinement regions is shown. We distinguish between three distinct regions, namely the interface refinement region for the residual points,
the interface refinement region for the points to fit the volume fraction field $\alpha$ and the nearfield refinement region for the residual points. The interface thickness $\delta_I$ predicted by the trained model can thus be controlled with $\delta_1$
and equals, depending on the quality of the fit, about $2\delta_1$. Furthermore, it is ensured that the residual points within the interface refinement region always measure a non-zero absolute gradient $|\nabla \alpha| \neq 0$. 

The total loss function $MSE_{total}$ for the inverse problem is composed of the losses associated with the fit of the interface, i.e. the volume fraction $MSE_\alpha$
plus the loss associated with the residual of the governing equations $MSE_f$. Additionally, in order to infer the unique solution that is given by the CFD, the neural network must be informed of the corresponding boundary conditions $MSE_{BC}$. 
\begin{align}
    MSE_{total} &= MSE_\alpha + MSE_{BC} + MSE_f \label{MSE_total_inverse} \\
    MSE_\alpha &= \frac{1}{N_\alpha} \sum_{i=1}^{N_\alpha} |\alpha^i - \hat{\alpha}(\mathbf{x}_\alpha^i, t_\alpha^i)|^2 \label{MSE_alpha}
\end{align}
Here, $\{\alpha^i, \mathbf{x}_\alpha^i, t_\alpha^i\}|_{i=1}^{N_\alpha}$ denotes the training data for the volume fraction. The computation of the boundary losses will be specified for each case individually.
Note that $MSE_f$ is computed as described by equations \eqref{eq:MSE_f} and \eqref{eq:weighting}.

\subsubsection{Drop in a shear flow}
\label{subsection:sheardrop}
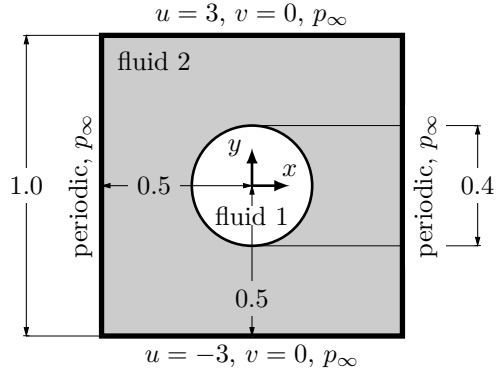
\begin{figure}[!b]
    \centering
    \begin{tikzpicture}
        \coordinate (A) at (0,0);
        \coordinate (B) at (2,2);
        \draw[line width=2pt, fill=black!20!white] (A) rectangle (4,4);
        \draw[line width=1pt, fill=white] (B) circle (0.8);
        \dimline[extension start length=1cm, extension end length=1cm,extension style={black}, label style={rotate=-90}] {(-1, 0)}{(-1, 4)}{1.0};
        \dimline[extension start length=-3cm, extension end length=-3cm,label style={rotate=-90}, extension style={black}] {(5,1.2)}{(5,2.8)}{0.4};
        \dimline[extension start length=0, extension end length=0, label style={rotate=-90, fill=black!20!white, near start}] {(2,0)}{(2,2)}{0.5};
        \dimline[extension start length=0, extension end length=0, label style={fill=black!20!white, xshift=-0.3cm}] {(0,2)}{(2,2)}{0.5};
        \node[rotate=90, above] at (0,2) {periodic, $p_\infty$};
        \node[rotate=90, below, fill=white, yshift=-0.028cm] at (4,2) {periodic, $p_\infty$};
        \node[above] at (2,4) {$u=3$, $v=0$, $p_\infty$}; 
        \node[below] at (2,0) {$u=-3$, $v=0$, $p_\infty$}; 
        \draw[->, line width=1pt] (B) -- ($(B)+(0.5,0.0)$) node[at end, above] {$x$};
        \draw[->, line width=1pt] (B) -- ($(B)+(0.0,0.5)$) node[at end, left] {$y$};
        \node[fill=white,inner sep=0pt,minimum size=1pt] at (2,1.6) {fluid 1};
        \node[below] at (0.7,3.9) {fluid 2}; 
    \end{tikzpicture}
    \caption{Schematic of the initial and boundary conditions for the drop in a shear flow.}
    \label{fig:sheardrop_setup}
\end{figure}

A liquid drop is immersed within another fluid and a shear flow is generated by moving boundaries. Viscous forces will cause the drop to deform, changing its
circular shape to an ellipsoidal shape. Capillary forces counteract this process, resulting in a steady state deformation at equilibrium. The final shape of the drop is described by the viscosity ratio
$\mu_b/\mu_d$ and the capillary number $Ca = \mu_bR\dot{\gamma}/\sigma$ \citep{Taylor1934}, where $R$ is the initial drop radius, $\dot{\gamma}$ is the shear rate, $\sigma$ is the surface tension coefficient
and the indices b and d indicate the bulk and the drop phase, respectively. This is a standard case studied in two-phase modeling \citep{Luo2015b,Hu2007a}. Figure \ref{fig:sheardrop_setup} shows the initial and boundary conditions for the present setup.
The two fluids are bound by the spatio-temporal domain $(x,y)\in[-0.5,0.5]\times[-0.5,0.5]$ and $t\in [0.0,0.54]$. The initial drop radius is $R=0.2$ and the surface tension coefficient is $\sigma=5$. 
The viscosity $\mu$ and density $\rho$ of both fluids is 1.0. The reference length scale $L_r=R$ is used for non-dimensionalization. All other reference quantities are 1.0.

In addition to the data on the interface, the neural network is informed about the fixed value north and south boundaries, where $u=3$, $v=0$ and $u=-3$, $v=0$, respectively, as well as the periodic east and west boundaries. Furthermore, an ambient pressure of $p_\infty$
is provided at the boundaries at $t=0$. Since the pressure field in incompressible flows is unique only up to a constant, the value of the ambient pressure may be chosen rather arbitrarily. For consistency 
with the exponential activation of the respective output node, we set the pressure field to be positive by a proper choice of $p_\infty$. Additionally, $p_\infty$ should not be chosen too large, as all labels should stay within the same order of magnitude, i.e. the 
labels should be normalized. With that in mind, an ambient pressure $p_\infty=10$ is prescribed.

The loss functions for the boundary conditions are given by
\begin{align*}
    MSE_{NS} &=  \frac{1}{N_{NS}} \sum_{i=1}^{N_{NS}} |u^i_{NS} - \hat{u}(\mathbf{x}^i_{NS},t^i_{EW})|^2 + |v^i_{NS} - \hat{v}(\mathbf{x}^i_{NS},t^i_{NS})|^2, \\
    MSE_{EW} &=  \frac{1}{N_{EW}} \sum_{i=1}^{N_{EW}} |\hat{u}(x^i_{E},y^i_{EW}, t^i_{EW}) - \hat{u}(x^i_{W}y^i_{EW}, t^i_{EW})|^2, \\
             & + |\hat{v}(x^i_{E},y^i_{EW}, t^i_{EW}) - \hat{v}(x^i_{W},y^i_{EW}, t^i_{EW})|^2 + |\hat{p}(x^i_{E},y^i_{EW}, t^i_{EW}) - \hat{p}(x^i_{W},y^i_{EW}, t^i_{EW})|^2, \\
    MSE_{NSEW} &= \frac{1}{N_{NSEW}} \sum_{i=1}^{N_{NSEW}} |p_{NSEW}^i - \hat{u}(\mathbf{x}^i_{NSEW},t^i_{NSEW})|^2, \\
    MSE_{BC} &= MSE_{NS} + MSE_{EW} + MS_{NSEW},
\end{align*}
where $\{u^i_{NS}, v^i_{NS}, \mathbf{x}^i_{NS}, t^i_{EW}\}|_{i=1}^{N_{NS}}$ denotes the training data for the north and south boundary condition and $\{x^i_{E},x^i_{W},y^i_{EW},t^i_{EW}\}|_{i=1}^{N_{EW}}$ represents the training points for the 
east and west boundary condition. The data $\{p_{NSEW}^i,\mathbf{x}^i_{NSEW},t^i_{NSEW}\}|_{i=1}^{N_{NSEW}}$ is used to enforce the ambient pressure at all boundaries at $t=0$.

\begin{figure}[!b]
    \input{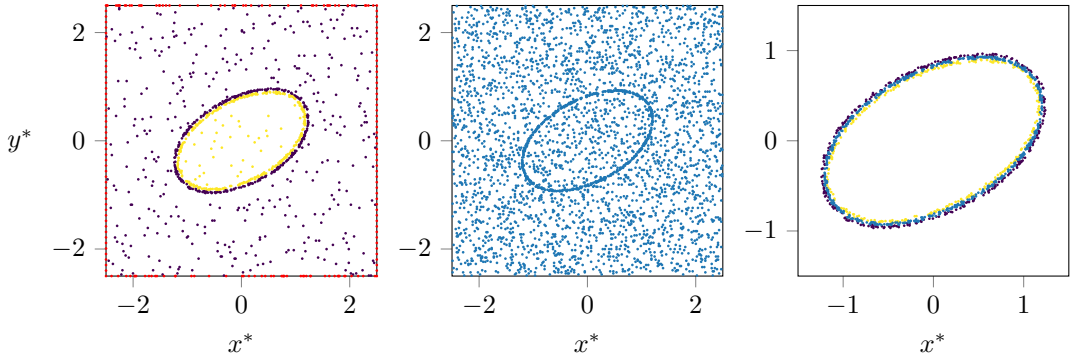}
    \caption{Point distribution for the drop in a shear flow at $t=0.54$. (Left) Points for the volume fraction $\alpha$ and for the boundary conditions.
    Yellow and violet marker colors indicate a volume fraction of 1 and 0, respectively. Red markers indicate points for the boundary conditions. (Middle) Residual points.
    (Right) Close up of the points that are used for the interface refinement. Notice the gap between the points corresponding to $\alpha=1$ and $\alpha=0$.
    This gap is a reference for the predicted interface thickness $\delta_I$ of the trained model.
    The residual points for interface refinement are placed within said gap, to ensure that these points always measure a non-zero absolute gradient $|\nabla \alpha|\neq 0$.}    
    \label{fig:sheardrop_point_distribution}
\end{figure}

For small Weber numbers, the loss of the momentum equation residual is orders of magnitude larger than all other losses due to the error between the surface tension term and the pressure gradient.
In particular, the gradient of the volume fraction $\nabla \alpha$ 
becomes larger the better the fit of the interface. Therefore, the loss has to be weighted accordingly. When using the mean squared error as loss function, we found that weighting the loss of the momentum equation residual with a factor
of $\mathcal{O}(-3)$ is suitable for cases with Weber numbers of $\mathcal{O}(-2)$. The Weber number here is $We=0.04$. Successively increasing the loss weights while monitoring the training history should be done to achieve the best possible results.
Here, the loss weights are $w_{f,m} = 1.0$, $w_{f,u} = \num{1e-3}$, $w_{f,v} = \num{1e-3}$, $w_{f,\alpha} = 1$. Note that $MSE_f$ is computed as described by equations \eqref{eq:MSE_f} and \eqref{eq:weighting}. A comparison of the
results for different loss weights and different activation functions is shown in \ref{section:A}.

\begin{figure}[!b]
    \centering
    \input{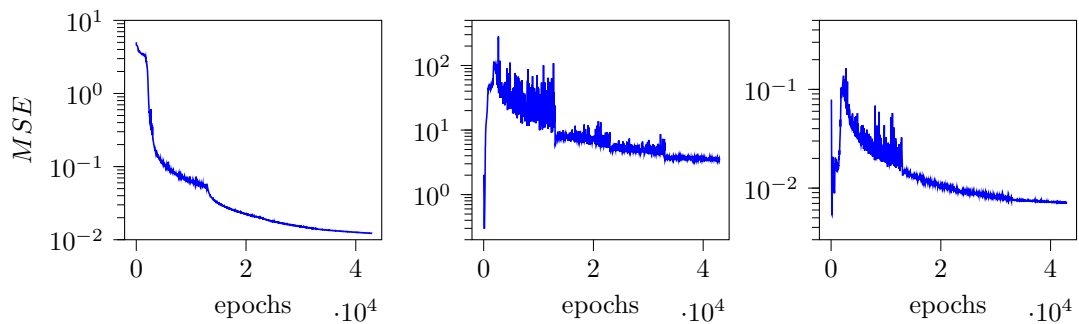}
    \caption{Training history of the drop in a shear flow. (Left) Losses associated with the volume fraction $MSE_\alpha$. (Middle) Losses associated with the $y-$momentum equation residual $MSE_{f,v}$.
    (Right) Losses associated with the residual of the continuity equation and the advection equation $MSE_{f,m}$ + $MSE_{f,\alpha}$.}
    \label{fig:sheardrop_loss_history}
\end{figure}

As for the training point distribution, the CFD solution provides data on a uniform 256$\times$256 grid with time snapshots linearly spaced with $\Delta t=0.005$ in the whole temporal domain $t\in[0.0,0.54]$. We use 21 of these time snapshots
to random uniformly distribute the spatial points for $\alpha$ and the residual points. In particular, the first 7 time snapshots are used, as it is important to provide data in the early stages when the shear flow is still developing. 
From there, 4, 5 and 5 time snapshots with a spacing of $\Delta t=0.015$, $\Delta t=0.03$ and $\Delta t=0.06$, respectively, are added to the set of time snapshots. At each of those, 
600 spatial points are scattered across the interface refinement region for $\alpha$, with $\delta_1=\num{4e-3}$ and $\delta_2=\num{8e-3}$ (see Figure \ref{fig:interface_refinement} for an illustration of the refinement regions) and
400 spatial points are used for the rest of the domain, hence a total of $N_\alpha=21000$ points. The interface refinement for the residual points is populated with 500 points, while for the rest of the domain 3000 points are used, resulting in $N_f=73500$.
No nearfield refinement is considered for this case. As for the boundary conditions, 20 time snapshots are random uniformly drawn within the respective bound for each boundary. Here, it is ensured that $t=0.0$ and $t=0.54$ are included in the set
of time snapshots and that the time snapshots for the east and west boundaries are equal, as this is necessary to enforce periodic behaviour. At each of the time snapshots, 50 spatial points are random uniformly drawn
for the north and south boundary and linearly spaced for the east and west boundary, hence $N_{NS} = 2000$ and $N_{EW} = 1000$. To enforce the reference pressure at all boundaries at $t=0$, the same points are used, thus $N_{NSEW} = 200$.
The distribution of the training points for the examplary time snapshot $t=0.54$ is depicted in Figure \ref{fig:sheardrop_point_distribution}.

\begin{table}[!t]
    \centering
    \setlength\arrayrulewidth{1pt}
    \begin{tabular}{lccccc}
        \hline
        epochs& 3000 & 10000 & 10000 & 10000 & 10000 \\
        learning rate & \num{5e-4} & \num{1e-4} & \num{5e-5} & \num{1e-5} & \num{5e-6}\\
        batches & 20 & 20 & 20 & 20 & 20\\
        \hline
    \end{tabular}
    \caption{Training hyper-parameters for the drop in a shear flow.}
    \label{table:sheardrop_hyperparameters}  
\end{table}

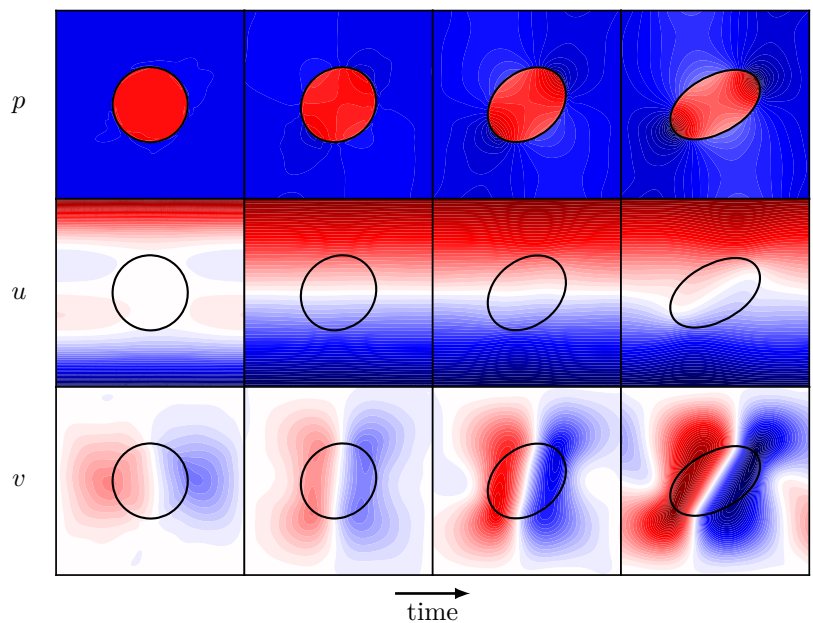
\begin{figure}[!b]
    \centering
    \begin{tikzpicture}
        \node at (0,0) {\includegraphics[scale=0.80]{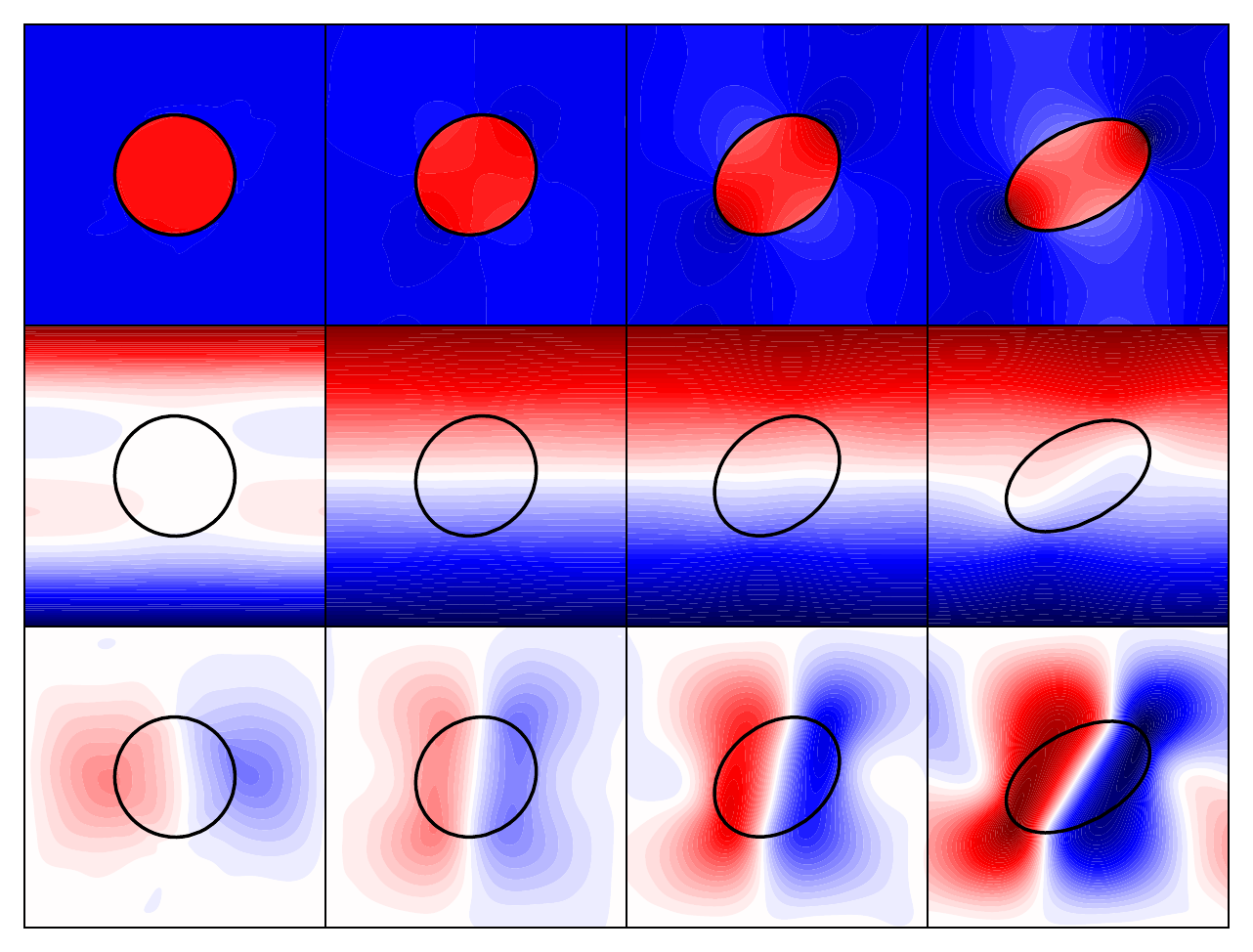}};
        \node at (-5.5,2.5) {$p$};
        \node at (-5.5,0.0) {$u$};
        \node at (-5.5,-2.5) {$v$};
        \draw[->, line width=1] (-0.5,-4) -- (0.5,-4) node[midway, below] {time};
    \end{tikzpicture}
    \caption{Qualitative time evolution of the drop in a shear flow for time snapshots $t=0, 0.05, 0.1, 0.5$ predicted by the neural network.
    the black lines indicate the interface.
    The colormap is ranging from maximum (red) to minimum (blue) value of the corresponding quantity within the entire spatio-temporal domain.}
    \label{fig:sheardrop_qualitative}
\end{figure}

The neural network that is used for this case maps $x,y,t \rightarrow u,v,p,\alpha$ and is composed of 7 hidden layers with 250 nodes each. The training hyper-parameters are listed in Table \ref{table:sheardrop_hyperparameters}.
In Figure \ref{fig:sheardrop_loss_history}, the training history is depicted. After about 2000 epochs with no significant reduction of any loss, the loss associated with the interface $MSE_\alpha$ aprubtly reduces, accompanied by a distinct jump 
of the losses associated with the residual of the PDE $MSE_{f,i}$ with $i=\{m,u,v,\alpha\}$. As explained earlier, the loss of the momentum equation residual increases by orders of magnitude due to the increasing gradient of the 
volume fraction $\nabla\alpha$. Here, the loss weights are crucial. If the momentum equation is not weighted accordingly, the optimizer will not find a better set of network parameters than what the initialization delivered.

\begin{figure}[!b]
    \centering
    \begin{tikzpicture}
        \node[anchor=south west] at (0,0) {\includegraphics[scale=0.5]{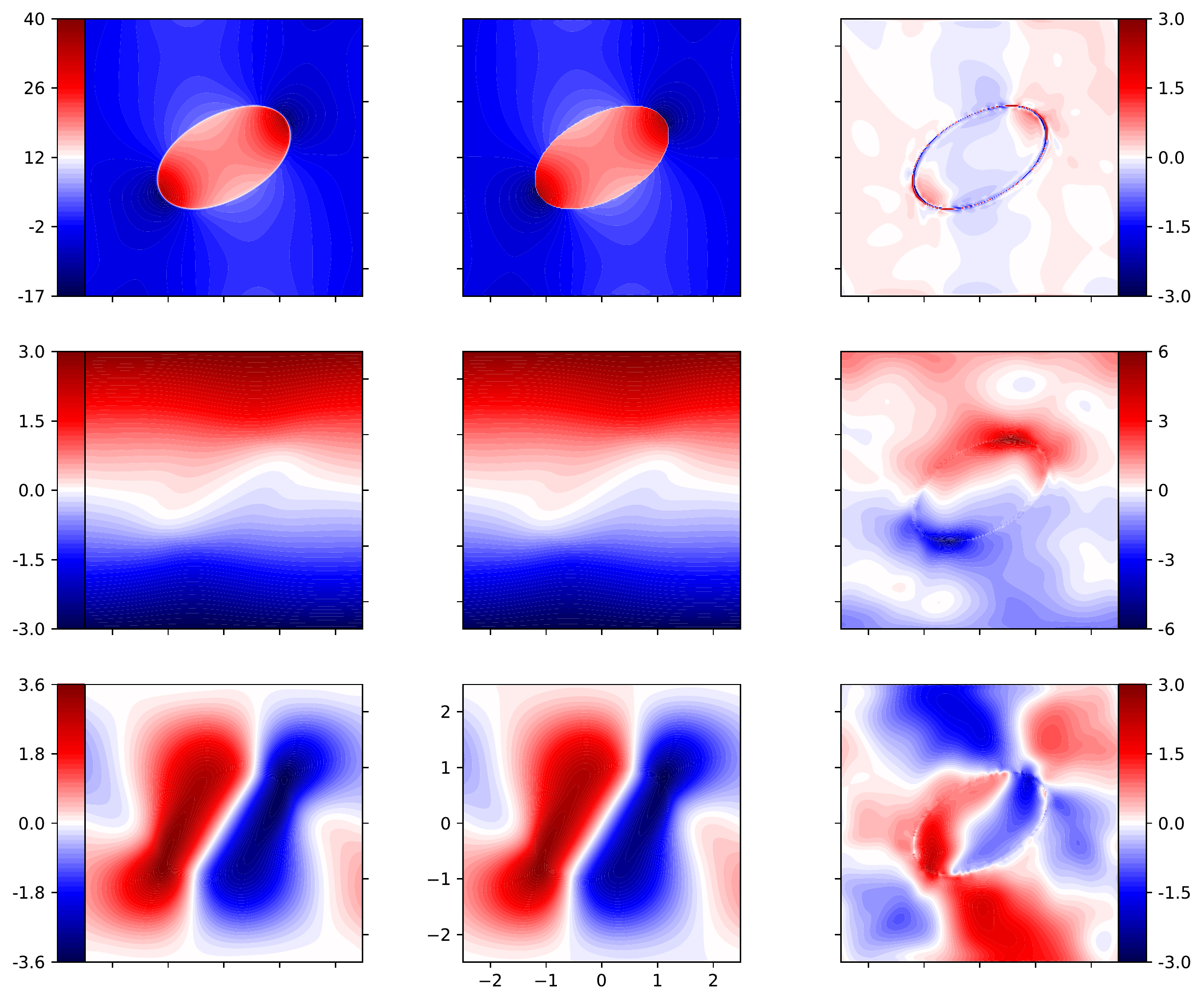}};
        \node at (6.35,0) {$x^*$};
        \node at (4.35,1.95) {$y^*$};
        \node at (2.3,10.7) {nn};
        \node at (6.35,10.7) {exact};
        \node at (10.3,10.7) {error};
        \node[rotate=90, align=center, anchor=mid] at (-0.2,2) {$v^* \cdot 10^{1}$};
        \node[rotate=90, align=center, anchor=mid] at (-0.2,5.45) {$u^*$};
        \node[rotate=90, align=center, anchor=mid] at (-0.2,8.95) {$p^*$};
        \node[rotate=90, align=center, anchor=mid] at (13,2) {$(v_{nn}^*-v_{exact}^*) \cdot 10^{2}$};
        \node[rotate=90, align=center, anchor=mid] at (13,5.4) {$(u_{nn}^*-u_{exact}^*) \cdot 10^{2}$};
        \node[rotate=90, align=center, anchor=mid] at (13,8.9) {$p_{nn}^*-p_{exact}^*$};
    \end{tikzpicture}
    \caption{Quantitative comparison of the prediction and the exact solution of the drop in a shear flow $t=0.5$.}
    \label{fig:sheardrop_quantitative}
\end{figure}

\begin{figure}[!t]
    \centering
    \input{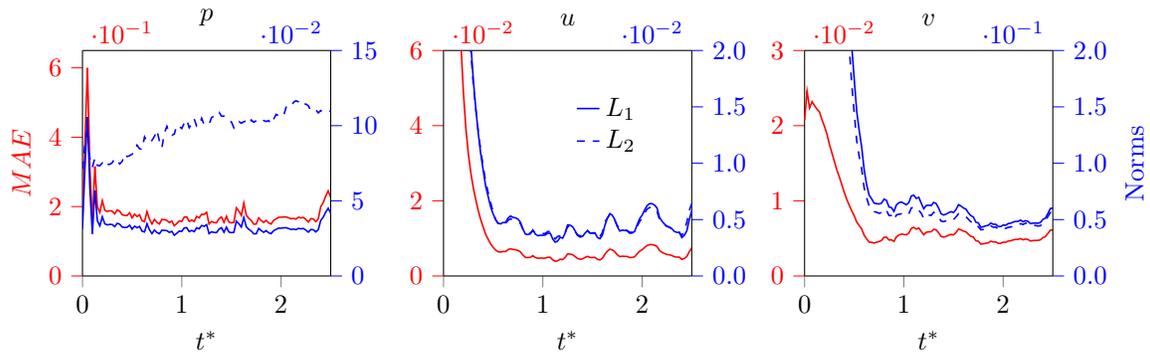}
    \vspace{-0.8cm}
    \caption{Mean absolute error and relative errors norms between prediction and exact solution of the drop in a shear flow.} 
    \label{fig:sheardrop_norms}
\end{figure}

The predicted qualitative time evolution is shown in Figure \ref{fig:sheardrop_qualitative}. Noticable are the non-zero velocities at $t=0$, which may be improved by increasing the amount of time snapshots at very early stages or introducing losses that prescribe initial conditions for $u$ and $v$. 
We compare the neural network prediction of the steady state at $t=0.5$ to the exact solution in Figure \ref{fig:sheardrop_quantitative}. The predicted and exact pressure field have been normalized by substracting the corresponding spatial average pressure from both fields,
since for incompressible flows the pressure is unique only up to a constant. In Figure \ref{fig:sheardrop_norms} the mean absolute error and the relative errors norms between prediction and exact solution over time are depicted.
Looking at the velocity, after high initial relative errors resulting from the fact that the exact velocity at $t=0$ is zero, the relative errors are about $0.5\%$ and $5\%$ for $u$ and $v$, respectively.
The relative $L_1$ and $L_2$ errors for the pressure are about $10\%$ and $4\%$, respectively. As for the quality of the interface fit, the predicted interface thickness $\delta_{I}$ is computed by measuring the orthogonal distance between the
isolines $\alpha=0.01$ and $\alpha=0.99$ in $x$ and $y$ direction and averaging the results for $t=0$. This results in $\delta_I=\num{8.674e-3}$, which is about $2\delta_1$.

\begin{table}[!b]
    \centering
    \setlength\arrayrulewidth{1pt}
    \begin{tabular}{lcccc}
        \cline{3-5}
        &&layers&scale factor&initial\\
        \hline
        \textit{adaptive 1} & $a$ & $1-7$ & 10 & 0.1 \\
        \multirow{2}{*}{\textit{adaptive 2}} & $a_1$ & $1-4$ & 20 & 0.05 \\
        & $a_2$ & $5-7$ & 10 & 0.1 \\

        \hline
    \end{tabular}
    \caption{Adaptive activation function configurations for the drop in a shear flow.}
    \label{table:sheardrop_ad_act}  
\end{table}

To study the influence of adaptive activation functions, the following two adaptive activation function configurations will be used. For the first one, referred to as \textit{adaptive 1}, a single adaptive activation
coefficient $a$ is introduced to each node of all hidden layers, with an initial value of $0.1$ and a scale factor of $10$. The second one, referred to as \textit{adaptive 2}, 
introduces two layer-wise adaptive activation coefficients $a_1$ and $a_2$ into each node of the hidden layers $1-4$ and $5-7$, respectively.  
We scale $a_2$ with a value of $10$ and initialize it with $0.1$. Since the first coefficient is exptected to be harder to optimize as it appears in earlier layers,
a scale factor of $20$ and an initial value of $0.05$ is used. This way, the optimization process is more sensitive towards $a_2$. Table \ref{table:sheardrop_ad_act} lists
both configurations.
\begin{figure}[!t]
    \centering
    \input{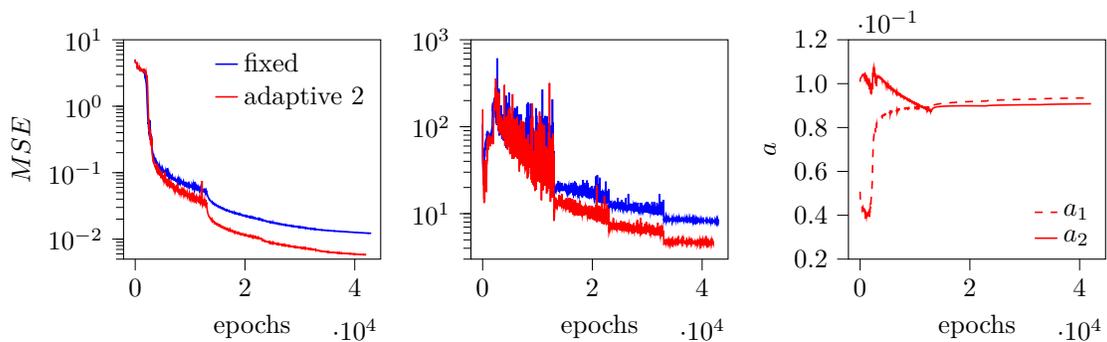}
    \caption{Training history of the drop in a shear flow for fixed and adaptive activation functions. (Left) Losses associated with the volume fraction $MSE_\alpha$. (Middle) Losses associated with the momentum equation residual $\sum_i MSE_{f,i}$, $i=\{u,v\}$.
    (Right) Adaptive activation coefficients.}
    \label{fig:sheardrop_loss_history_ad_act}
\end{figure}

The training history for the models \textit{fixed} and \textit{adaptive 2} is illustrated in Figure \ref{fig:sheardrop_loss_history_ad_act}. At around $\mathrm{epoch}=1750$, when $MSE_\alpha$ reduces
aprubtly, the coefficient $a_1$ increases significantly. At this point, $a_2$ also shows a small jump towards larger values, however, subsequently reduces again.
\begin{table}[!t]
    \centering
    \setlength\arrayrulewidth{1pt}
    \begin{tabular}{lcccccc}
        \hline
        &$MSE_\alpha$ & $MSE_{BC}$ & $MSE_{f,m}$ & $MSE_{f,u}$ & $MSE_{f,v}$ & $MSE_{f,\alpha}$ \\
        \hline
        \textit{fixed} & \num{1.22e-2} & \num{3.03e-4} & \num{4.13e-2} & \num{4.63e+0} & \num{3.49e+0} & \num{2.98e-3}\\
        \textit{adaptive 1}& \num{1.20e-2} & \num{7.09e-4} & \num{6.13e-3} & \num{8.89e+0} & \num{4.80e+0}& \num{3.73e-3}\\
        \textit{adaptive 2}& \num{5.81e-3} & \num{1.36e-4} & \num{1.71e-3} & \num{2.42e+0} & \num{2.10e+0}& \num{2.13e-3}\\
        \hline
    \end{tabular}
    \caption{Comparison of the final losses of the drop in a shear flow for fixed and adaptive activation functions.}
    \label{table:sheadrop_losses_ad_act}  
\end{table}
Only after the second learning rate reduction at $\mathrm{epoch}=13000$, $a_2$ increases successively with a very small rate. In Table \ref{table:sheadrop_losses_ad_act}, the final losses for fixed
and adaptive activation functions are listed. While \textit{adaptive 1} shows increased final losses, the configuration \textit{adaptive 2} significantly reduces all losses. An evaluation of the errors between prediction and exact solution
for the models \textit{fixed} and \textit{adaptive 2} reveals that there is a signifcant reduction of the initial relative errors of $u$ and $v$ (compare Figure \ref{fig:sheardrop_norms}), i.e. the inference of the initial zero velocity is improved. For the rest of the temporal domain though,
no major improvements regarding the relative errors are observed. The average computational cost per epoch increases by about $30\%$ when using the configurations \textit{adaptive 1} and \textit{adaptive 2} compared to \textit{fixed}.

\subsubsection{Oscillating droplet}
\label{subsection:oscillatingdrop}
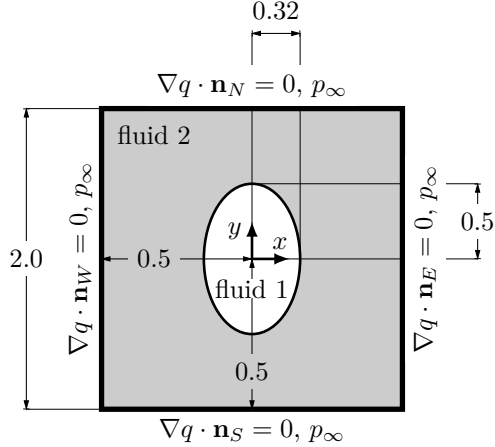
\begin{figure}[!b]
    \centering
    \begin{tikzpicture}
        \coordinate (A) at (0,0);
        \coordinate (H) at (4,4);
        \coordinate (B) at (2,2);
    
        \draw[line width=2pt, fill=black!20!white] (A) rectangle (H);
        \draw[line width=1pt, fill=white] (B) ellipse (0.64 and 1);
        \dimline[extension start length=1cm, extension end length=1cm,extension style={black}, label style={rotate=-90}] {(-1, 0)}{(-1, 4)}{2.0};
        \dimline[extension start length=-3cm, extension end length=-3cm,label style={rotate=-90}, extension style={black}] {(5,2)}{(5,3)}{0.5};
        \dimline[extension start length=3cm, extension end length=3cm, extension style={black}, label style={yshift=0.3cm}] {(2,5)}{(2.64,5)}{0.32};
        \dimline[extension start length=0, extension end length=0, label style={rotate=-90, fill=black!20!white, near start}] {(2,0)}{(2,2)}{0.5};
        \dimline[extension start length=0, extension end length=0, label style={fill=black!20!white, xshift=-0.3cm}] {(0,2)}{(2,2)}{0.5};
        \node[rotate=90, above] at (0,2) {$\nabla q \cdot \textbf{n}_W=0$, $p_\infty$};
        \node[rotate=90, below, fill=white] at (4.029,2) {$\nabla q \cdot \textbf{n}_E=0$, $p_\infty$};
        \node[above, fill=white] at (2,4.029) {$\nabla q \cdot \textbf{n}_N=0$, $p_\infty$}; 
        \node[below] at (2,0) {$\nabla q \cdot \textbf{n}_S=0$, $p_\infty$};
        \draw[->, line width=1pt] (B) -- ($(B)+(0.5,0.0)$) node[near end, above] {$x$};
        \draw[->, line width=1pt] (B) -- ($(B)+(0.0,0.5)$) node[near end, left] {$y$};
        \node[fill=white,inner sep=0pt,minimum size=1pt] at (2,1.6) {fluid 1};
        \node[below] at (0.7,3.9) {fluid 2};
    \end{tikzpicture}
    \caption{Schematic of the initial and boundary conditions for the oscillating drop. Here, $q$ represents an arbitrary quantitiy.}
    \label{fig:oscillatingdrop_setup}
\end{figure}
An ellipsoidal drop is immersed into another fluid. Surface tension forces will drive an oscillation that is associated with a transfer between potential and kinetic energy.
The oscillation period is given by $T=2\pi\sqrt{(\rho_b+\rho_d)R^3/6\sigma}$ \citep{Strutt1879}, where indices $b$ and $d$ indicate the bulk and drop phase, respectively. The radius $R$ is the effective circle radius.
This is a further standard test case studied in two-phase modeling \citep{Luo2015b, Garrick2017b}. In Figure \ref{fig:oscillatingdrop_setup}, a schematic of the initial and boundary conditions for the present configuration is shown. The two fluids are bound by the spatio-temporal domain
$(x,y)\in[-1.0,1.0]\times[-1.0,1.0]$ and $t\in[0.0,0.45]$. The initial drop is an ellipse with a major and minor axis of 0.5 and 0.32, respectively. This results in an effective circle radius of $R=0.4$.
The surface tension coefficient is $\sigma=5$. Both fluids have the same density $\rho_{1/2}=1$ and viscosity $\mu_{1/2}=0.01$. The reference length scale $L_r=R$ is used for non-dimensionalization. All other reference quantities are 1.0.
\begin{figure}[!b]
    \centering
    \input{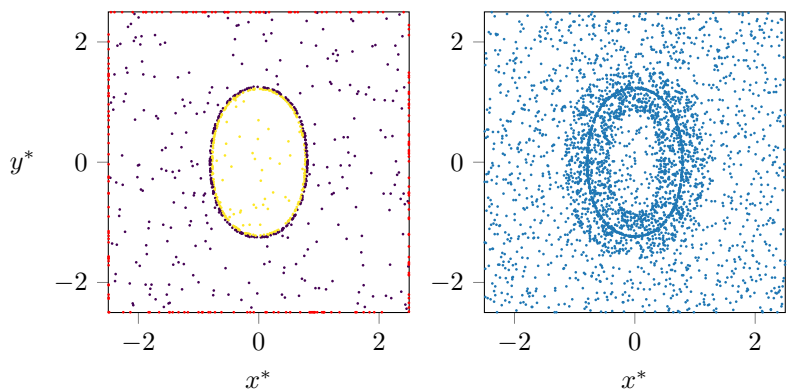}
    \caption{Point distribution for the oscillating drop at $t=0.0$. (Left) Points for the volume fraction $\alpha$ and the boundary conditions.
    Yellow and violet marker colors indicate a volume fraction of 1 and 0, respectively. Red markers indicate points for the boundary conditions.
    (Right) Residual points.}    
    \label{fig:oscillatingdrop_points}
\end{figure}
The boundary conditions are zero gradient for $u$ and $v$ and an ambient pressure of $p_\infty=10$. Furthermore, the neural network
is given actual measurements of the velocities $u$ and $v$ at the boundaries. While this is not necessary to infer a unique solution, these measurements have been found to
significantly improve the inference of the velocity field. A comparison of the predictions with and without measurements for $u$ and $v$ at the boundaries is shown in \ref{section:B}.

The loss functions are computed as follows:
\begin{align*}
    MSE_{NS} &=  \frac{1}{N_{NS}} \sum_{i=1}^{N_{NS}} |\hat{u}_y(\mathbf{x}^i_{NS},t^i_{NS})|^2 + |\hat{v}_y(\mathbf{x}^i_{NS},t^i_{NS})|^2, \\
    MSE_{EW} &=  \frac{1}{N_{EW}} \sum_{i=1}^{N_{EW}} |\hat{u}_x(\mathbf{x}^i_{EW},t^i_{EW})|^2 + |\hat{v}_x(\mathbf{x}^i_{EW},t^i_{EW})|^2, \\
    MSE_{NSEW} &= \frac{1}{N_{NSEW}} \sum_{i=1}^{N_{NSEW}} |u_{NSEW}^i - \hat{u}(\mathbf{x}^i_{NSEW},t^i_{NSEW})|^2 + |v_{NSEW}^i - \hat{v}(\mathbf{x}^i_{NSEW},t^i_{NSEW})|^2,\\
    & + |p_{NSEW}^i - \hat{p}(\mathbf{x}^i_{NSEW},t^i_{NSEW})|^2, \\
    MSE_{BC} &= MSE_{NS} + MSE_{EW} + MSE_{NSEW},
\end{align*}
where $\{\mathbf{x}^i_{NS}, t^i_{EW}\}|_{i=1}^{N_{NS}}$ denotes the points for the north and south zero gradient boundary condition and $\{\mathbf{x}^i_{EW},t^i_{EW}\}|_{i=1}^{N_{EW}}$ represents the points for the 
east and west zero gradient boundary condition. The data $\{u_{NSEW}^i, v_{NSEW}^i, p_{NSEW}^i, \mathbf{x}^i_{NSEW},t^i_{NSEW}\}|_{i=1}^{N_{NSEW}}$ is used to enforce the measurements for $u$ and $v$ as well as the ambient pressure.
For this case, the loss weights are $w_{f,m} = 1.0$, $w_{f,u} = \num{5e-3}$, $w_{f,v} = 5e-3$, $w_{f,\alpha} = 1$. Note that $MSE_f$ is computed as described by equations \eqref{eq:MSE_f} and \eqref{eq:weighting}.

\begin{table}[!b]
    \centering
    \setlength\arrayrulewidth{1pt}
    \begin{tabular}{lccccc}
        \hline
        epochs& 2000 & 5000 & 5000 & 7000 & 10000 \\
        learning rate & \num{1e-4} & \num{5e-5} & \num{1e-5} & \num{5e-6} & \num{1e-6}\\
        batches & 20 & 20 & 20 & 20 & 20\\
        \hline
    \end{tabular}
    \caption{Training hyper-parameters for the oscillating drop.}
    \label{table:oscillatingdrop_hyperparameters}  
\end{table}

\begin{figure}[!b]
    \centering
    \input{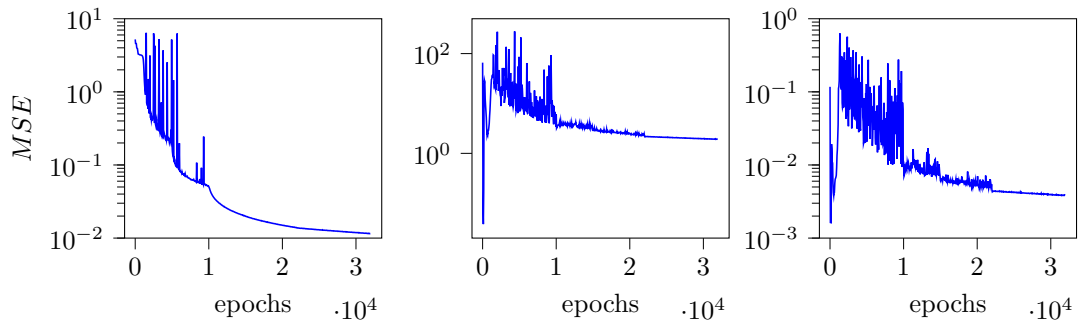}
    \caption{Training history for the oscillating drop. (Left) Losses associated with the volume fraction $MSE_\alpha$. (Middle) Losses associated with the momentum equation residual $\sum_i MSE_{f,i}$, $i=\{u,v\}$.
    (Right) Losses associated with the residual of the continuity equation and the advection equation $\sum_i MSE_{f,i}$, $i=\{m,\alpha\}$.}
    \label{fig:oscillatingdrop_history}
\end{figure}

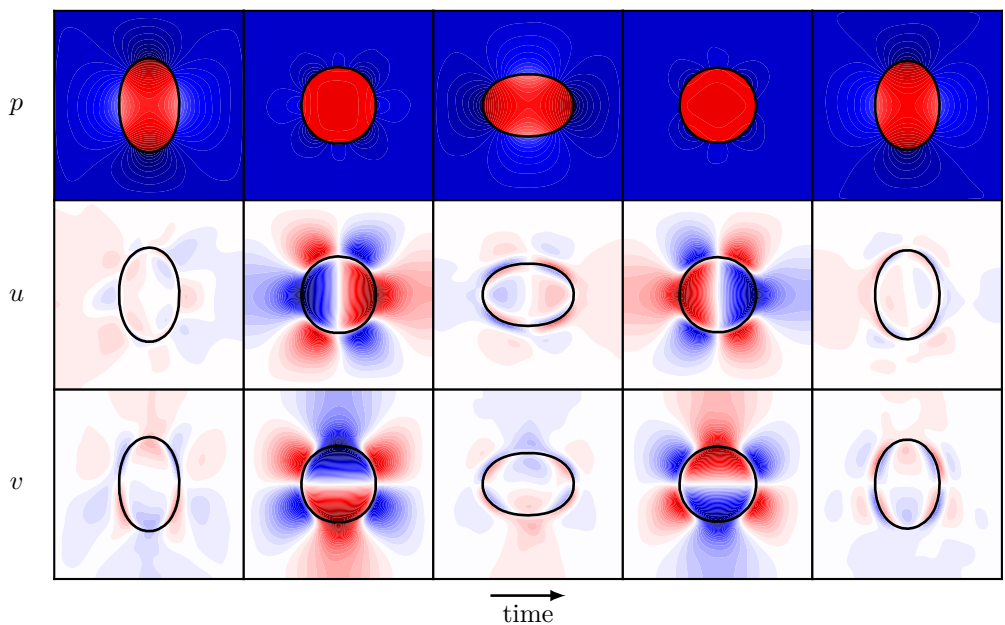
\begin{figure}[!b]
    \centering
    \begin{tikzpicture}
        \node at (0,0) {\includegraphics[scale=1.0]{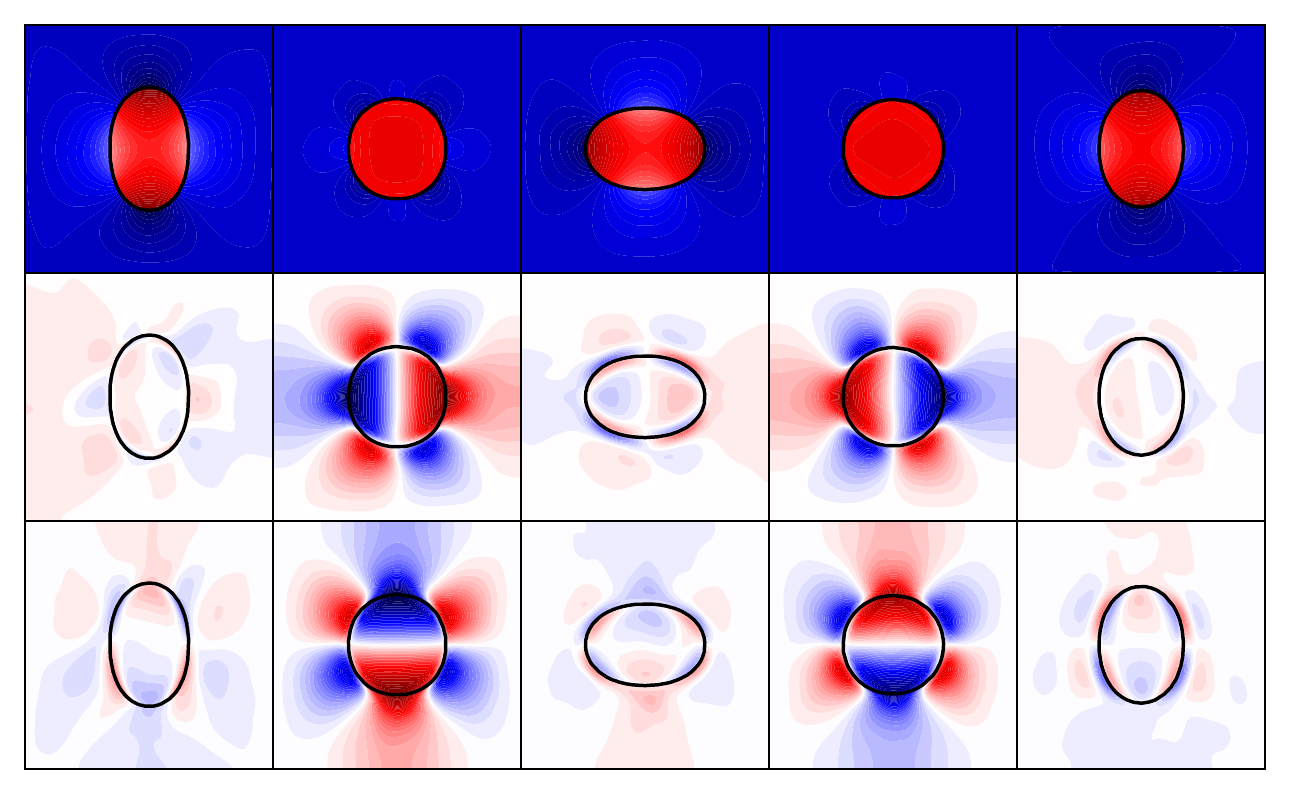}};
        \node at (-6.8,2.5) {$p$};
        \node at (-6.8,0) {$u$};
        \node at (-6.8,-2.5) {$v$};
        \draw[->, line width=1] (-0.5,-4) -- (0.5,-4) node[midway, below] {time};
    \end{tikzpicture}
    \caption{Qualitative time evolution of the oscillating drop for time snapshots $t=0.00, 0.11, 0.22, 0.33, 0.44$ predicted by the neural network. The black lines indicate the interface. 
    The colormap is ranging from maximum (red) to minimum (blue) value of the corresponding quantity within the entire spatio-temporal domain.}
    \label{fig:oscillatingdrop_qualitative}
\end{figure}

The CFD solution provides data on a uniform 512$\times$512 grid in linearly spaced time snapshots with $\Delta t = 0.005$ for two entire oscillation periods. Here, we infer one oscillation period. For the present configuration the oscillation period is $T=0.41$.
The temporal domain however is $t\in[0.0,0.45]$, as a few time snapshots are added before and after to better capture the still interface at maximum potential energy correlated with an ellipsodial shape.
In particular, we use 46 linearly spaced time snapshots, i.e. $\Delta t=0.01$, to random uniformly distribute the spatial points for the volume fraction $\alpha$ and the residual points.
At each time snapshot, the interface refinement region for $\alpha$ is populated with 500 points, with $\delta_1 = \num{4e-3}$ and $\delta_2 = \num{8e-3}$. For the rest of the domain, 300 points are used, resulting in $N_\alpha=36800$.
The inferface and nearfield refinement region for the residual points are filled with 600 and 1500 points, respectively, using $\delta_3 = \num{1.5e-2}$ (see Figure \ref{fig:interface_refinement} for illustration of the refinement regions).
The rest of the domain holds 1500 points, which leads to $N_f = 165600$. As for the boundary conditions, 20 time snapshots are random uniformly drawn from the time snapshots provided by the CFD within the respective bound. Here, it is ensured that $t=0.0$ and $t=0.45$ are in the set of time snapshots.
At each of these time snapshots, 50 spatial points are random uniformly distributed at each boundary. Thus the corresponding amount of samples are $N_{NS} = 2000$, $N_{EW} = 2000$ and $N_{NSEW} = 4000$. Note that the set of points for the minimization of $MSE_{NSEW}$ consists of the set of points
for the minimization of $MSE_{NS}$ and $MSE_{EW}$. In Figure \ref{fig:oscillatingdrop_points}, the point distribution for the examplary time snapshot $t=0.0$ is shown.
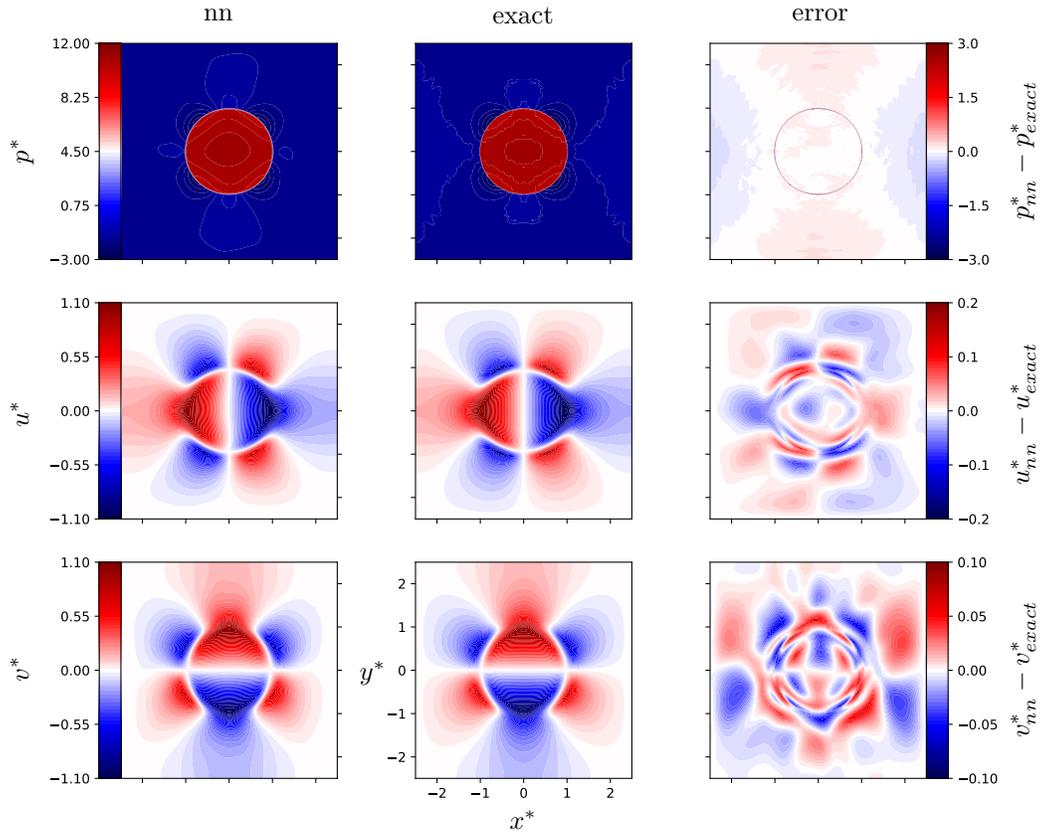
\begin{figure}[!t]
    \centering
    \begin{tikzpicture} 
        \node[anchor=south west] at (-0.2,0) {\includegraphics[scale=0.50]{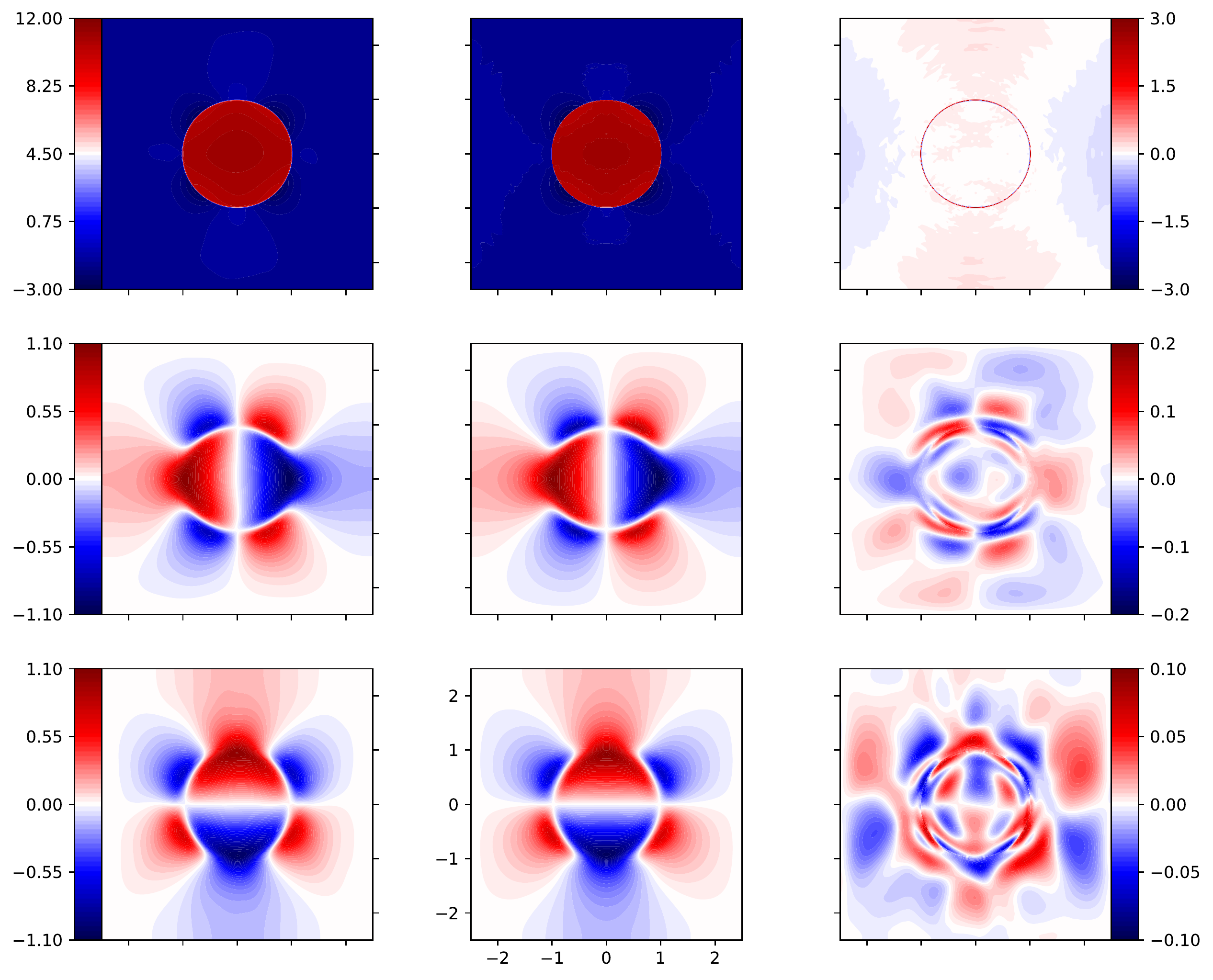}};
        \node at (6.35,0) {$x^*$};
        \node at (4.35,1.95) {$y^*$};
        \node at (2.3,10.7) {nn};
        \node at (6.35,10.7) {exact};
        \node at (10.3,10.7) {error};
        \node[rotate=90, align=center, anchor=mid] at (-0.3,2) {$v^*$};
        \node[rotate=90, align=center, anchor=mid] at (-0.3,5.4) {$u^*$};
        \node[rotate=90, align=center, anchor=mid] at (-0.3,8.9) {$p^*$};
        \node[rotate=90, align=center, anchor=mid] at (13,2) {$v_{nn}^*-v_{exact}^*$};
        \node[rotate=90, align=center, anchor=mid] at (13,5.4) {$u_{nn}^*-u_{exact}^*$};
        \node[rotate=90, align=center, anchor=mid] at (13,8.9) {$p_{nn}^*-p_{exact}^*$};
    \end{tikzpicture}
    \caption{Quantitative comparison of the prediction and the exact solution of the oscillating drop at $t=0.33$.}
    \label{fig:oscillatingdrop_quantitative}
\end{figure}

\begin{figure}[!t]
    \centering
    \input{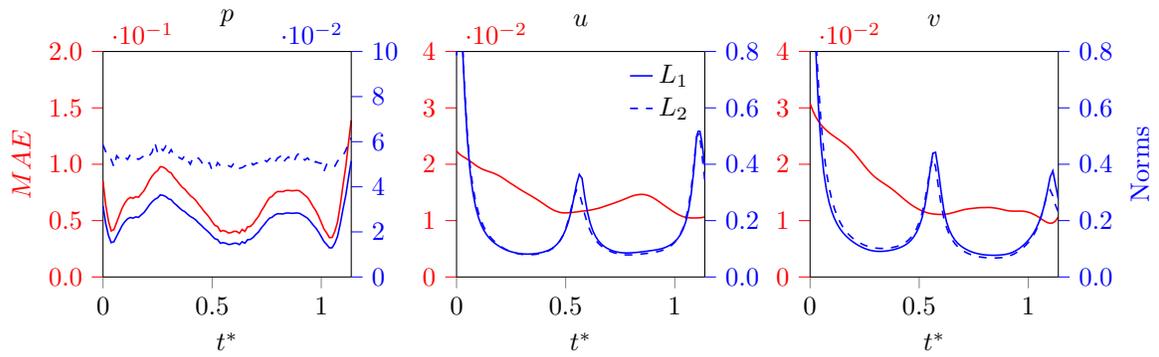}
    \vspace{-0.8cm}
    \caption{Mean absolute error and relative error norms between prediction and exact solution for the oscillating drop.}
    \label{fig:oscillatingdrop_norms}
\end{figure}

\begin{figure}[!t]
    \centering
\begin{tikzpicture}

\begin{axis}[
width=6cm,
height=5cm,
legend cell align={left},
legend style={draw=none, anchor=north east, at={(0.99,0.99)}},
tick align=outside,
tick pos=left,
xtick style={color=black},
ytick style={color=black},
xlabel=$t^*$,
ylabel=$E/E_{max}$
]
\addplot [semithick, blue]
table {%
0 0.0138746276497841
0.0124975 0.0123480008915067
0.0249925 0.0228735916316509
0.0374875 0.045104518532753
0.049985 0.0783744230866432
0.06248 0.121728740632534
0.0749775 0.174038931727409
0.0875025 0.233957156538963
0.1 0.299656927585602
0.1125 0.369587182998657
0.1249975 0.441894948482513
0.1374975 0.514956176280975
0.1499975 0.586843967437744
0.1624975 0.656211256980896
0.1749975 0.721604406833649
0.1874975 0.781749546527863
0.1999975 0.835493326187134
0.2124975 0.882126986980438
0.225 0.920985877513885
0.2375 0.951571464538574
0.25 0.973269760608673
0.2625 0.985940754413605
0.2750025 0.989706635475159
0.2875025 0.984160900115967
0.3000025 0.969412684440613
0.3125025 0.945685744285583
0.3250025 0.91333532333374
0.3375025 0.872883141040802
0.3500025 0.824476897716522
0.3625025 0.769161522388458
0.3750025 0.70817494392395
0.3875 0.642845332622528
0.4 0.574253618717194
0.4124975 0.504283547401428
0.4249975 0.434618532657623
0.437495 0.366572201251984
0.449995 0.301855772733688
0.4624925 0.241605266928673
0.47499 0.18699224293232
0.4874875 0.138929843902588
0.499985 0.0981392338871956
0.5124825 0.0651589184999466
0.52498 0.0403943173587322
0.5375025 0.0240463968366385
0.55 0.0162722691893578
0.562495 0.0169835314154625
0.5749925 0.02599654532969
0.58749 0.0429935939610004
0.599985 0.0675090849399567
0.6124825 0.0989477261900902
0.6249825 0.136597514152527
0.63748 0.179550752043724
0.6499775 0.226862147450447
0.662505 0.277572304010391
0.6750025 0.330297082662582
0.6875025 0.384006053209305
0.7 0.437430679798126
0.7125 0.489488065242767
0.725 0.538948774337769
0.7374975 0.584812104701996
0.7499975 0.626088500022888
0.7624975 0.661773502826691
0.774995 0.691150188446045
0.787495 0.7135249376297
0.799995 0.728508472442627
0.812495 0.735750675201416
0.824995 0.735139131546021
0.8374925 0.726892471313477
0.8499925 0.711213767528534
0.8624925 0.688597857952118
0.8749925 0.659665048122406
0.8874925 0.625022113323212
0.8999925 0.585613191127777
0.9124925 0.542135775089264
0.924995 0.495651870965958
0.937495 0.446999192237854
0.949995 0.39728257060051
0.962495 0.34729990363121
0.974995 0.298096179962158
0.987495 0.250475436449051
0.999995 0.205259338021278
1.012495 0.163285419344902
1.024995 0.125246912240982
1.037495 0.0917658060789108
1.049995 0.0634880438446999
1.0624925 0.0408974625170231
1.0749925 0.0243630800396204
1.08749 0.0140762403607368
1.0999875 0.00997487083077431
1.112485 0.0117109687998891
1.124985 0.0186268202960491
1.1374825 0.0297786556184292
};
\addplot [semithick, blue, mark=o, mark size=2, mark repeat=3, only marks]
table {%
0 0.0191110340856146
0.0124975 0.0188975041978865
0.0249925 0.0300638265125697
0.0374875 0.0524526013517488
0.049985 0.0858276258872078
0.06248 0.129179359499885
0.0749775 0.181676823012472
0.0875025 0.241972163164177
0.1 0.308238892223973
0.1125 0.379188041963739
0.1249975 0.45239654312547
0.1374975 0.526030724069286
0.1499975 0.5982123144041
0.1624975 0.668143362053997
0.1749975 0.732813177359556
0.1874975 0.792643666471917
0.1999975 0.84590023475284
0.2124975 0.892137062006836
0.225 0.930520891751707
0.2375 0.959323982991333
0.25 0.982173917425348
0.2625 0.994888717507986
0.2750025 1
0.2875025 0.994717886458135
0.3000025 0.982166332212462
0.3125025 0.958756009082941
0.3250025 0.927346347227071
0.3375025 0.887323967054933
0.3500025 0.838995185290667
0.3625025 0.784048371987985
0.3750025 0.722438200839291
0.3875 0.656945543740588
0.4 0.587885809804718
0.4124975 0.517201647716041
0.4249975 0.446936638116296
0.437495 0.378192851025524
0.449995 0.31226625722103
0.4624925 0.250885527954535
0.47499 0.195209251142735
0.4874875 0.145926549140176
0.499985 0.104031620753459
0.5124825 0.0699387357606479
0.52498 0.0442171653918385
0.5375025 0.0269836183911695
0.55 0.0184612752381219
0.562495 0.0184608245209546
0.5749925 0.0267817172611891
0.58749 0.0431359980147524
0.599985 0.0669971981832253
0.6124825 0.0977389862726774
0.6249825 0.134881041170315
0.63748 0.177346732215338
0.6499775 0.22432304738146
0.662505 0.274771984535114
0.6750025 0.32745498179631
0.6875025 0.381563949897627
0.7 0.435688507149876
0.7125 0.488454452902047
0.725 0.538840730246774
0.7374975 0.585618649751747
0.7499975 0.628784494217617
0.7624975 0.665795043959481
0.774995 0.696279066843746
0.787495 0.719735744028193
0.799995 0.736139294834998
0.812495 0.744383592575042
0.824995 0.745207805070975
0.8374925 0.737514116451753
0.8499925 0.722251360458233
0.8624925 0.700367561009195
0.8749925 0.671837561977588
0.8874925 0.637109414395495
0.8999925 0.597412345526173
0.9124925 0.553596385195879
0.924995 0.50673224789841
0.937495 0.457736649730367
0.949995 0.407225630789489
0.962495 0.356654325803294
0.974995 0.306731123031472
0.987495 0.258229312371651
0.999995 0.212129106500724
1.012495 0.169411405085748
1.024995 0.130612179065111
1.037495 0.0963487358757379
1.049995 0.0674588409983283
1.0624925 0.0442152945248784
1.0749925 0.0271706904430305
1.08749 0.0164650067692132
1.0999875 0.012260566669323
1.112485 0.0145455625283503
1.124985 0.0231486182507336
1.1374825 0.0377537623276446
};
\end{axis}

\end{tikzpicture}
    \caption{Comparison of the predicted and exact kinetic energy $E=\int\int \alpha\rho\mathbf{u}\cdot\mathbf{u} \mathrm{dxdy}$ of the oscillating drop with $E_{max}$ being the maximum energy of the exact solution.
    The prediction and the exact solution are represented by a line and markers, respectively.}
    \label{fig:oscillatingdrop_energy}
\end{figure}
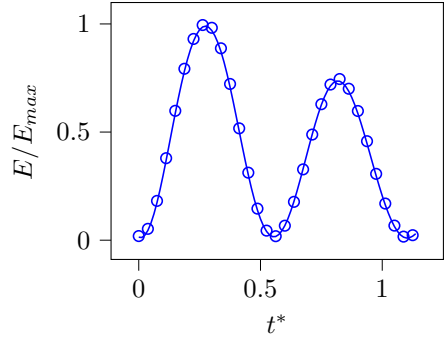

The deep neural network used for this case maps $x,y,t\rightarrow u,v,p,\alpha$ and consists of 8 hidden layers each composed of 300 nodes.
The training hyper-parameters are listed in table \ref{table:oscillatingdrop_hyperparameters}. The training history is shown in Figure \ref{fig:oscillatingdrop_history}.
The characteristics of the curves are similar to the previous case. After about 1000 epochs, $MSE_\alpha$ reduces significantly, accompanied with a distinct jump toward larger values of
$MSE_{f,i}$, $i=\{m,u,v,a\}$. Subsequently, all losses reduce successively. Noticable here is that for the period of the first two learning rates, i.e. first 10000 epochs, $MSE_\alpha$
is significantly noisier compared to the drop in a shear flow. 

In Figure \ref{fig:oscillatingdrop_qualitative}, the predicted qualitative time evolution is shown. In particular, the time snapshots corresponding to the minima and maxima
of the kinetic energy of the drop are displayed. Noticable are the non-zero velocities at the times of maximum potential energy, i.e. when the drop has an ellipsoidal shape. As described earlier, the temporal 
domain was already extended by a few time snapshots before and after the start and end of the oscillation period for this very reason. The examplary time snapshot $t=0.33$ of the prediction 
is compared to the exact solution in Figure \ref{fig:oscillatingdrop_quantitative}. The pressure fields of both the prediction and the exact solution have been normalized by
subtracting the corresponding mean pressure from the pressure field. This is necessary since the pressure in incompressible flows is unique only up to a constant.

\begin{table}[!b]
    \centering
    \setlength\arrayrulewidth{1pt}
    \begin{tabular}{lcccc}
        \cline{3-5}
        &&layers&scale factor&initial\\
        \hline
        \textit{adaptive 1} & $a$ & $1-8$ & 10 & 0.1 \\
        \multirow{2}{*}{\textit{adaptive 2}} & $a_1$ & $1-4$ & 20 & 0.05 \\
        & $a_2$ & $5-8$ & 10 & 0.1 \\
        \hline
    \end{tabular}
    \caption{Configuration of the adaptive activation coefficients for the drop in a shear flow and the rising bubble.}
    \label{table:oscillatingdrop_ad_act}  
\end{table}

Figure \ref{fig:oscillatingdrop_norms} displays the mean absolute error and the relative error norms between the prediction and the exact solution over time. High relative errors 
in $u$ and $v$ are observed in particular at the start and the end of oscillation period but also at around $t=0.22$. These are the times of maximum potential energy, i.e. the exact velocities are almost zero,
which is the reason for the high relative errors. In the transition regions, where the drop is going from an ellipsoidal to a circular shape, the relative error norms for the velocities range between $8\%-10\%$.
As for the pressure, the relative $L_1$ and $L_2$ errors are about $3\%$ and $5\%$, respectively. The predicted and exact kinetic energy of the drop over time is compared in Figure \ref{fig:oscillatingdrop_energy}.
The largest relative errors of the kinetic energy are observed at the peaks where it is about $3\%$.

\begin{table}[!t]
    \centering
    \setlength\arrayrulewidth{1pt}
    \begin{tabular}{lcccccc}
        \hline
        &$MSE_\alpha$ & $MSE_{BC}$ & $MSE_{f,m}$ & $MSE_{f,u}$ & $MSE_{f,v}$ & $MSE_{f,\alpha}$ \\
        \hline
        \textit{fixed} & \num{1.15e-2} & \num{2.34e-4} & \num{2.75e-3} & \num{9.99e-1} & \num{9.21e-1} & \num{1.05e-3}\\
        \textit{adaptive 1}& \num{1.13e-2} & \num{2.04e-4} & \num{3.14e-3} & \num{8.79e-1} & \num{8.77e-1}& \num{1.07e-3}\\
        \textit{adaptive 2}& \num{8.00e-3} & \num{1.92e-4} & \num{2.59e-3} & \num{7.37e-1} & \num{6.99e-1}& \num{7.37e-4}\\
        \hline
    \end{tabular}
    \caption{Comparison of the final losses of the oscillating drop for fixed and adaptive activation functions.}
    \label{table:oscillatingdrop_losses_ad_act}  
\end{table}


Table \ref{table:oscillatingdrop_ad_act} lists the configurations that are used to study the influence of adaptive activation functions for this test case.
The according final losses for \textit{fixed} and \textit{adaptive} are shown, in Table \ref{table:oscillatingdrop_losses_ad_act}.
The average computational time per epoch increases by about $30\%$ for both adaptive configurations. 


\subsubsection{Rising bubble}
\label{subsection:risingbubble}
\begin{figure}[!t]
    \input{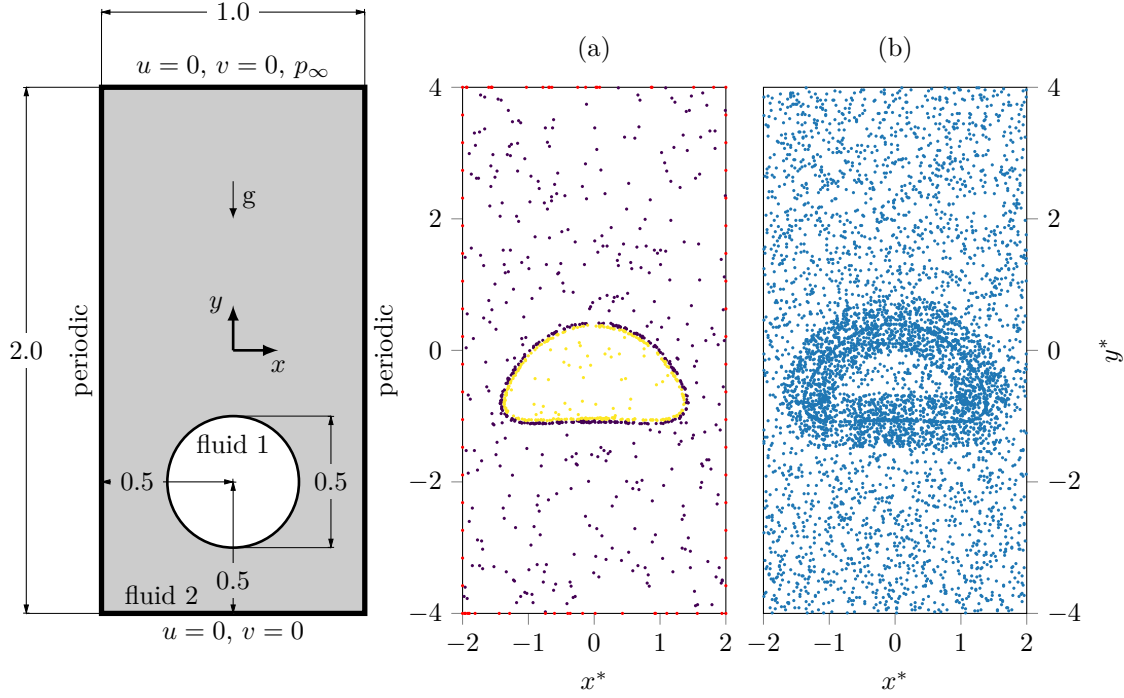}
    \caption{(Left) Schematic of the initial and boundary conditions for the rising bubble. (Right) Point distribution for the rising bubble at $t=1.5$. (a) Points for the volume fraction $\alpha$ and the boundary conditions.
    Yellow and violet marker colors indicate a volume fraction of 1 and 0, respectively. Red markers indicate points for the boundary conditions. (b) Residual points.}
    \label{fig:risingbubble_setup}
\end{figure}

A bubble is immersed into another, heavier fluid. Buoyancy causes the bubble to rise and subsequently undergo moderate shape deformations. This test case 
was studied by \cite{Hysing2009b} as a benchmark case to compare various two-phase solver. Figure \ref{fig:risingbubble_setup} shows a schematic of the initial 
and boundary conditions on the left. The two fluids are bound by the spatio-temporal domain $(x,y)\in[-0.5,0.5]\times[-1.0,1.0]$ and $t\in[0.0,3.0]$. The initial bubble radius is $R=0.25$. 
The densities and viscosities are $\rho_1=100$, $\rho_2=1000$, $\mu_1=1$ and $\mu_2=10$, respectively. The surface tension coefficient is $\sigma=24.5$ and the gravitational acceleration 
is $\mathbf{g}=[0,-0.98]^T$. The reference density $\rho_r=\rho_2$ and reference length scale $L_r=R$ is used for non-dimensionalization, while all other reference quantities are 1.0. At the north and south boundary, the no slip condition $u=0$ and $v=0$
is imposed, while the east and west boundaries are periodic. Furthermore, an ambient pressure of $\textcolor{red}{p_\infty^*}=1$ is prescribed at the north boundary.

The losses associated with the boundary conditions are computed as follows:
\begin{align*}
    MSE_{NS} &=  \frac{1}{N_{NS}} \sum_{i=1}^{N_{NS}} |u^i_{NS} - \hat{u}(\mathbf{x}^i_{NS},t^i_{EW})|^2 + |v^i_{NS} - \hat{v}(\mathbf{x}^i_{NS},t^i_{NS})|^2 \\
    MSE_{EW} &=  \frac{1}{N_{EW}} \sum_{i=1}^{N_{EW}} |\hat{u}(x^i_{E},y^i_{EW}, t^i_{EW}) - \hat{u}(x^i_{W}y^i_{EW}, t^i_{EW})|^2 \\
             & + |\hat{v}(x^i_{E},y^i_{EW}, t^i_{EW}) - \hat{v}(x^i_{W},y^i_{EW}, t^i_{EW})|^2 + |\hat{p}(x^i_{E},y^i_{EW}, t^i_{EW}) - \hat{p}(x^i_{W},y^i_{EW}, t^i_{EW})|^2 \\
    MSE_{N} &= \frac{1}{N_{N}} \sum_{i=1}^{N_{N}} |p_N^i - \hat{u}(\mathbf{x}^i_{N},t^i_{N})|^2 \\
    MSE_{BC} &= MSE_{NS} + MSE_{EW} + MSE_{N}
\end{align*}
Here, $\{u^i_{NS},v^i_{NS},\mathbf{x}^i_{NS},t^i_{NS})\}|_{i=1}^{N_{NS}}$ denotes the training data for the north/south no slip boundary, $\{x^i_E,x^i_W,y_{EW}^i,t_{EW}^i\}|_{i=1}^{N_{EW}}$ are the points for the east/west periodic boundary
and $\{p_N^i,\mathbf{x}^i_{N},t^i_{N})\}|_{i=1}^{N_N}$ represents the data to enforce the ambient pressure at the north boundary. The loss weights are $w_{f,m} = 1.0$, $w_{f,u} = 10$, $w_{f,v} = 10$, $w_{f,\alpha} = 1$. Note that
unlike in the previous cases, $MSE_{f,u}$ and $MSE_{f,v}$ are weighted with the same order of magnitude as all other losses. This is because the Weber number here is $\mathcal{O}(2)$ larger than previously, i.e. $We=10.2$. 

\begin{table}[!b]
    \centering
    \setlength\arrayrulewidth{1pt}
    \begin{tabular}{lccccc}
        \hline
        epochs& 5000 & 5000 & 5000 & 5000 & 5000 \\
        learning rate & \num{1e-4} & \num{5e-5} & \num{1e-5} & \num{5e-6}& \num{1e-6}\\
        batches & 20 & 20 & 20 & 20 & 20\\
        \hline
    \end{tabular}
    \caption{Training hyper-parameters for the rising bubble.}
    \label{table:risingbubble_hyperparameters}  
\end{table}


The CFD solution provides data on a uniform 256$\times$512 grid in linearly spaced time snapshots for the whole temporal domain with $\Delta t=0.02$, of which 29 will be used to random uniformly distribute the spatial
points for the volume fraction $\alpha$ and for the residual points. With $t=0$ being the first time snapshot, 2 and 24 time snapshots with $\Delta t=0.3$ and $\Delta t=0.1$, respectively, are added to the set of time snapshots.
Furthermore, $t=0.02$ and $t=0.04$ are included, to provide more information of the still interface in the beginning for better inference of the zero velocity at $t=0$. At each time snapshot, the interface refinement region for the volume fraction $\alpha$ is populated with 
500 spatial points, while for the rest of the domain 400 spatial points are used, resulting in $N_\alpha=26100$. The interface and nearfield refinement region for the residual points is filled with $500$ and $2500$ points, respectively. The rest of the domain contains $4000$ residual points, which leads to $N_f=203000$. 
Here, the width of the refinement regions are $\delta_1 = \num{4e-3}$, $\delta_2=\num{8e-3}$ and $\delta_3=\num{1e-2}$. As for the boundary conditions, 20 time snapshots are random uniformly drawn within the respective bound for each boundary. Here we ensure that $t=0.0$ and $t=3.0$ are included in the set of time snapshots. Furthermore, it is ensured 
that the time snapshots for the east and west boundary are equal, since this is necessary to enforce periodicity. At each of those time snapshots, 20 spatial points are random uniformly drawn for the north/south boundary and linearly spaced for the east/west boundary. This results in $N_{NS} = 800$ and $N_{EW} = 400$.
To enforce the ambient pressure at the north boundary, the same points are used, thus $N_N=400$. The point distribution for a time snapshot $t=1.5$ is depicted in Figure \ref{fig:risingbubble_setup} on the right. 

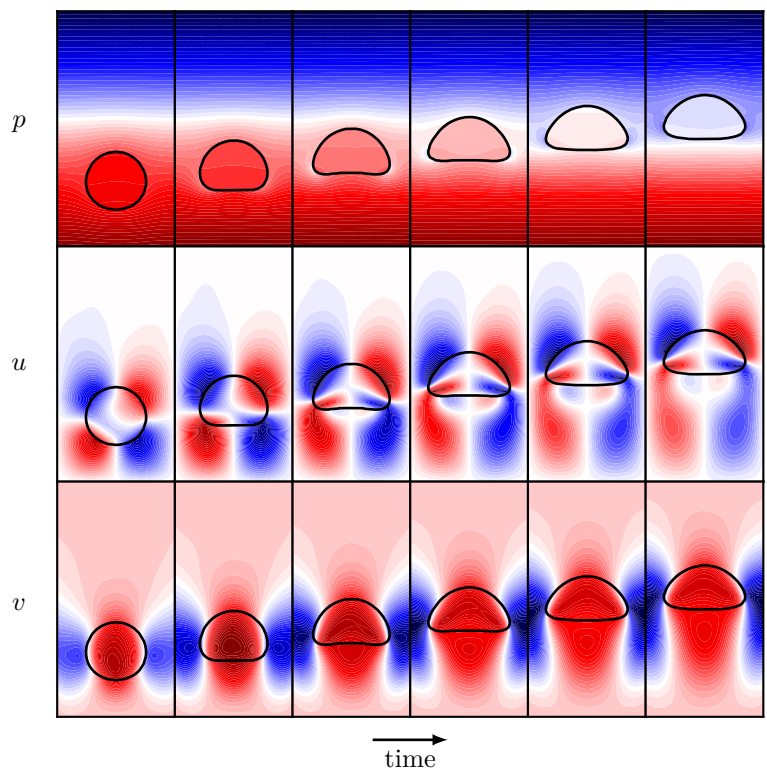
\begin{figure}[!b]
    \centering
    \begin{tikzpicture}
        \node at (0,0) {\includegraphics[scale=1.0]{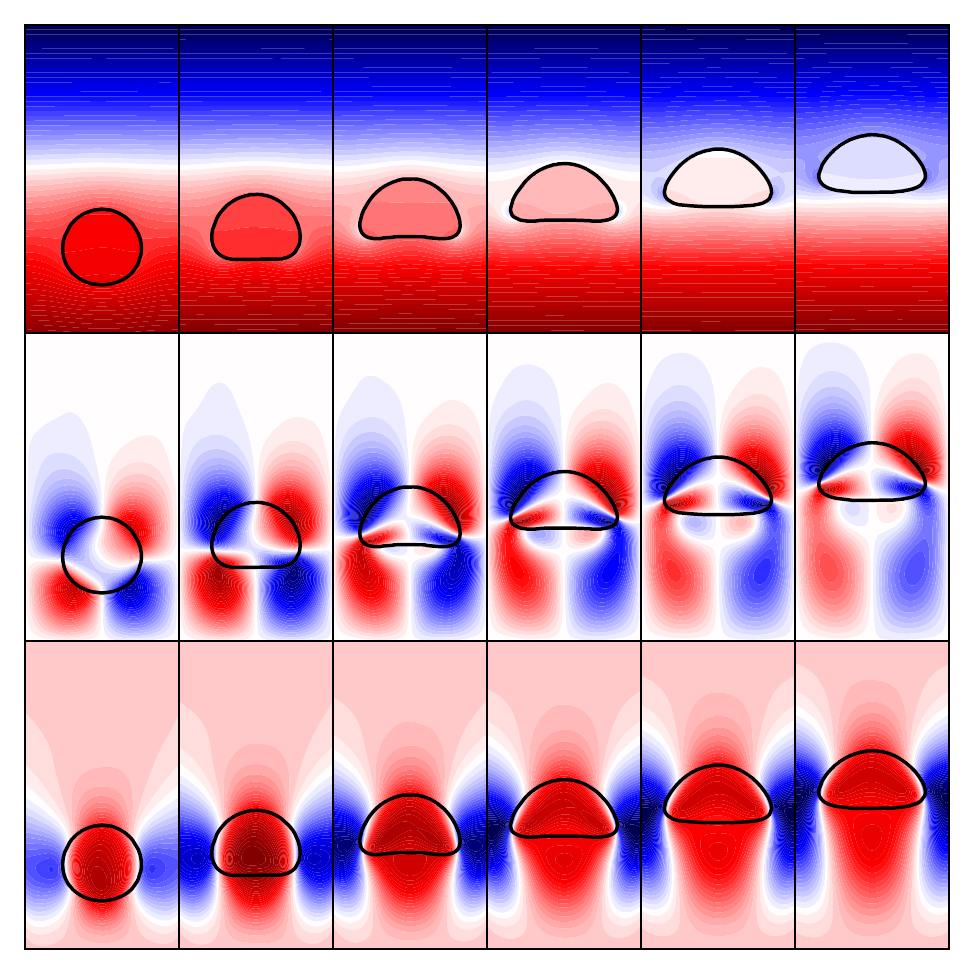}};
        \node at (-5.2,3.2) {$p$};
        \node at (-5.2,0) {$u$};
        \node at (-5.2,-3.2) {$v$};
        \draw[->, line width=1] (-0.5,-5) -- (0.5,-5) node[midway, below] {time};
    \end{tikzpicture}
    \caption{Qualitative time evolution of the rising bubble for time snapshots $t=0.5, 1.0, 1.5, 2.0, 2.5, 3.0$ predicted by the neural network. 
    The black line indicates the interface. The colormap is ranging from maximum (red) to minimum (blue) value of the corresponding quantity within the entire spatio-temporal domain.}
    \label{fig:risingbubble_qualitative}
\end{figure}

\begin{figure}[!t]
    \centering
    \begin{tikzpicture}
        \node[anchor=south west] at (0,0) {\includegraphics[scale=0.6]{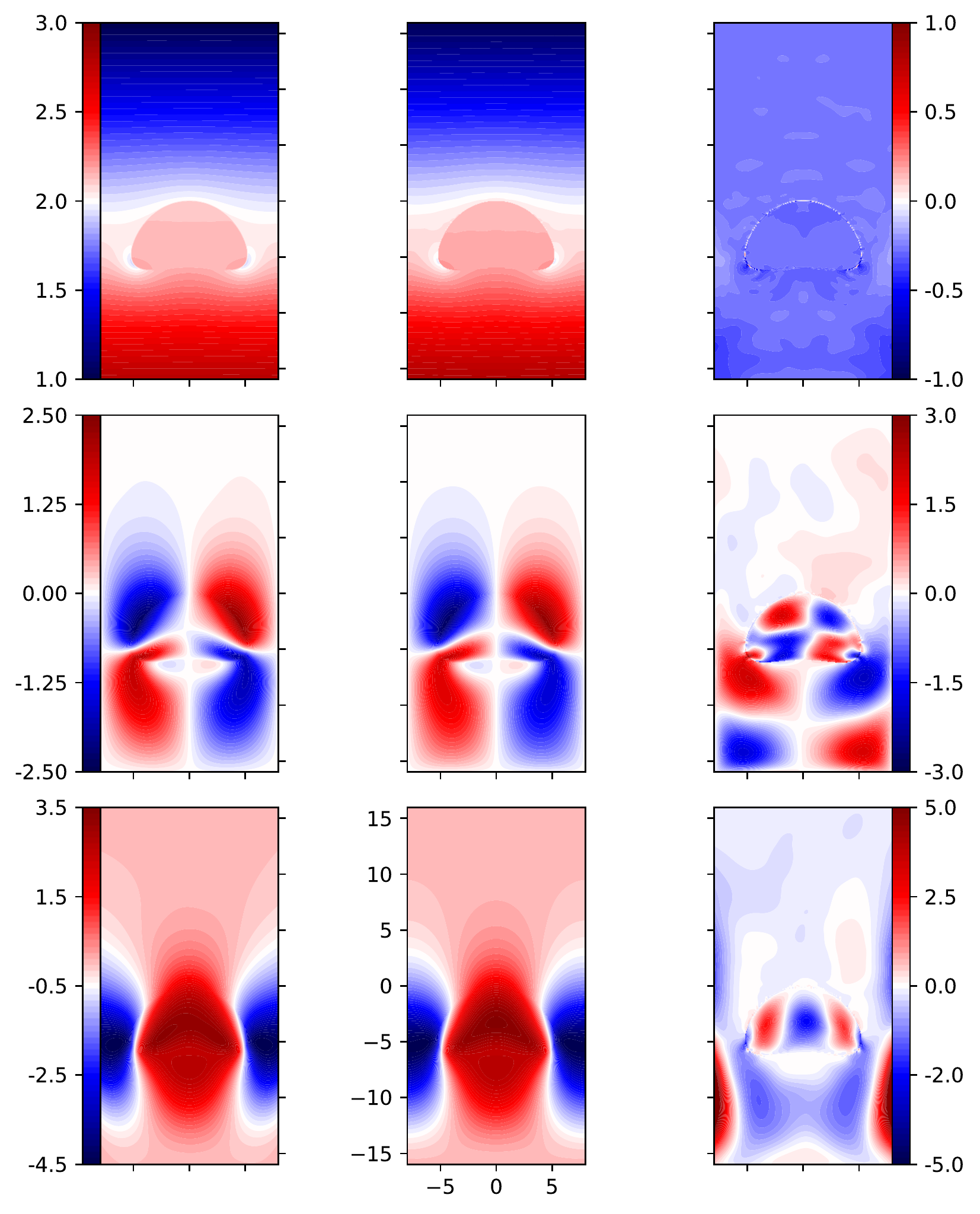}};
        \node at (5.25,0) {$x^*$};
        \node at (3.65,2.45) {$y^*$};
        \node at (2,12.6) {nn};
        \node at (5.22,12.6) {exact};
        \node at (8.42,12.6) {error};
        \node[rotate=90] at (0,2.5) {$v^* \cdot 10^{1}$};
        \node[rotate=90] at (0,6.5) {$u^* \cdot 10^{1}$};
        \node[rotate=90] at (0,10.5) {$p^*$};
        \node[rotate=90] at (10.5,2.5) {$(v_{nn}^*-v_{exact}^*) \cdot 10^{2}$};
        \node[rotate=90] at (10.5,6.5) {$(u_{nn}^*-u_{exact}^*) \cdot 10^{2}$};
        \node[rotate=90] at (10.5,10.5) {$(p_{nn}^*-p_{exact}^* )\cdot 10^{1}$};
    \end{tikzpicture}
    \caption{Quantitative comparison of the prediction and the exact solution of the rising bubble at $t=1.5$.}
    \label{fig:risingbubble_quantitative}
\end{figure}
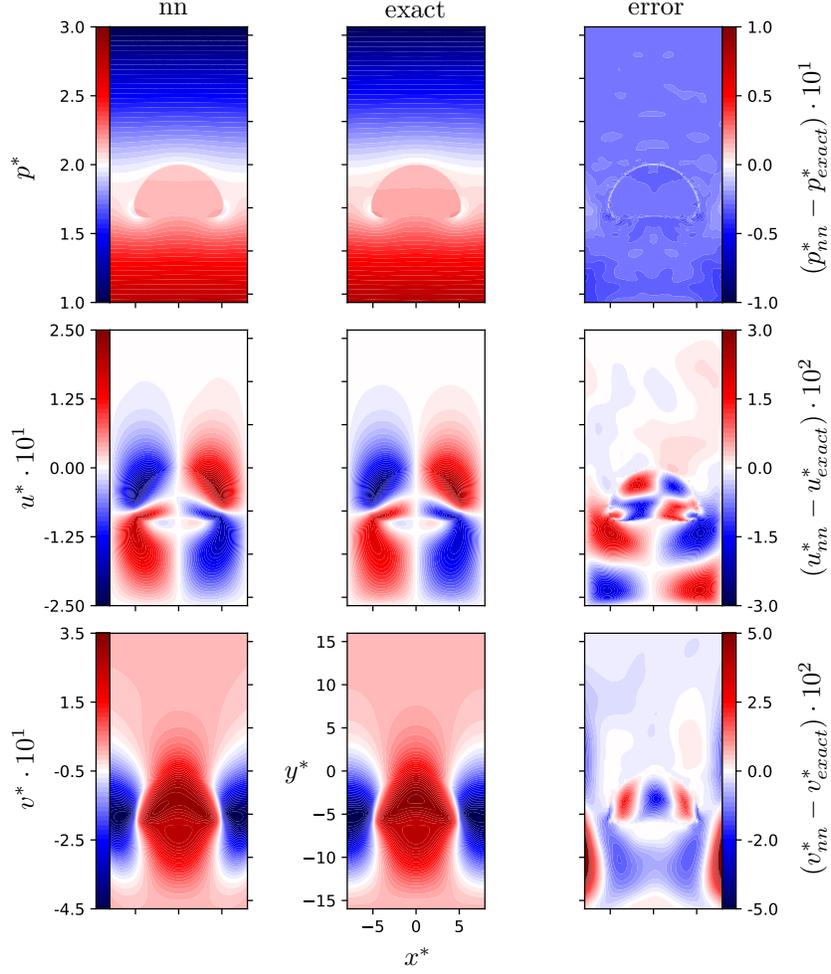

\begin{figure}[!t]
    \centering
    \input{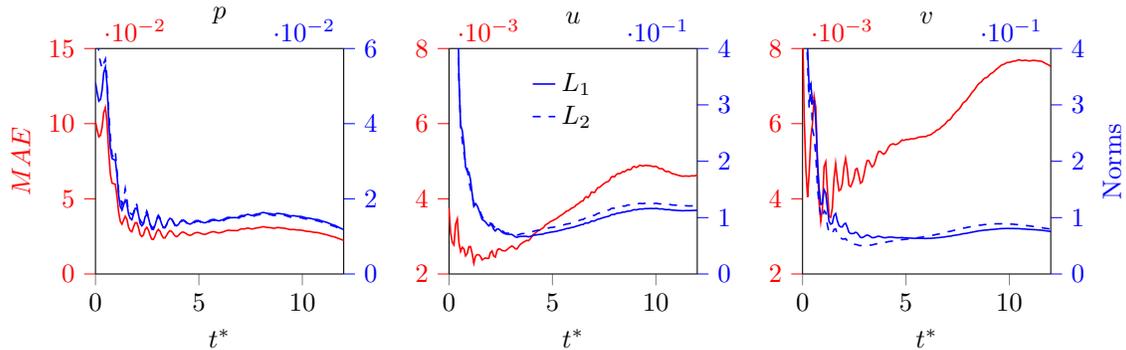}
    \caption{Mean absolute error and relative error norms between prediction and exact solution of the rising bubble.} 
    \label{fig:risingbubble_norms}
\end{figure}

For this case a neural network mapping $x,y,t \rightarrow u,v,p,\alpha$ consisting of 8 hidden layers with 400 nodes each is used.
The training hyper-parameters are listed in Table \ref{table:risingbubble_hyperparameters}. The predicted qualitative time evolution is shown in Figure \ref{fig:risingbubble_qualitative}.
Noticable is the symmetry-breaking error in $u$ for early times within the bubble. This may be reduced by distributing more residual points within the bubble for these times.
In Figure \ref{fig:risingbubble_quantitative}, a comparison between the prediction and the exact solution for an examplary time snapshots $t=1.5$ is depicted. To compare the predicted and exact pressure field,
they are normalized by subtracting the corresponding mean pressure. The errors between prediction and exact solution 
over time are displayed in Figure \ref{fig:risingbubble_norms}. The high initial relative errors of $u$ and $v$ are due to the fact that the exact velocity at $t=0$ is 0. Moving forward in time, the relative 
errors for $u$ and $v$ are about $10\%$ and $8\%$, respectively. As for the pressure, the high initial relative errors are explained by the initialization of the CFD. Here, only a linear hydrostatic pressure profile within 
the bulk fluid has been prescribed, while the initial pressure within the drop is constant. The average relative errors of the pressure are about $2\%$.




\begin{table}[!t]
    \centering
    \setlength\arrayrulewidth{1pt}
    \begin{tabular}{lcccccc}
        \hline
        &$MSE_\alpha$ & $MSE_{BC}$ & $MSE_{f,m}$ & $MSE_{f,u}$ & $MSE_{f,v}$ & $MSE_{f,\alpha}$ \\
        \hline
        \textit{fixed} & \num{2.81e-3} & \num{6.94e-5} & \num{1.17e-3} & \num{1.45e-4} & \num{1.39e-4} & \num{2.88e-3}\\
        \textit{adaptive 1}& \num{2.60e-3} & \num{4.23e-5} & \num{1.20e-3} & \num{1.40e-4} & \num{1.38e-4}& \num{2.21e-3}\\
        \textit{adaptive 2}& \num{2.12e-3} & \num{3.40e-5} & \num{1.04e-3} & \num{1.24e-4} & \num{1.18e-4}& \num{2.21e-3}\\
        \hline
    \end{tabular}
    \caption{Comparison of the final losses of the rising bubble for fixed and adaptive activation functions.}
    \label{table:risingbubble_losses_ad_act}  
\end{table}

Table \ref{table:risingbubble_losses_ad_act} shows the final losses for fixed and adaptive activation functions. The configurations \textit{adaptive 1} and \textit{2} are the same as in the previous case (compare \textcolor{red}{Table} \ref{table:oscillatingdrop_ad_act}).
Both adaptive configurations result in a reduction of all final losses. \textit{Adaptive 2} in particular improves the relative $L_1$ error to the exact solution, reducing the error for $u$, $v$, and $p$ on average over time by $10\%$, $8.8\%$ and $0.3\%$.
The computational cost per epoch increases by about $28\%$ for both configurations.

\section{Conclusion}
\label{section:4}
Physics-informed neural networks give rise to a new approach for the quantification of flow fields by combining available data of auxiliary variables 
with the underlying physical laws. This way, the velocity and pressure of entire flow fields may be inferred, for which a direct measurement is usually impracticle.
We extend previous work, where this method is applied to incompressible \citep{Raissi2019a} and compressible \citep{Mao2020b} single-phase flows, 
to incompressible two-phase flows using a Volume-of-fluid approach. The auxiliary variable here is the interface position, i.e. the
volume fraction field, motivated by experimentally accessible data of the interface position using a combination of optical techniques and image processing \citep{Murai2001,Takamasa2003,Murai2006,Poletaev2020}.

We investigate three classical test cases in two-phase modeling: A drop in a shear flow, an oscillating drop and a rising bubble.
The training data is generated with ALPACA \citep{Winter2019,Kaiser2019,Kaiser2019a}, a compressible two-phase solver that capures the interface using the
Level-set method. An effective strategy of distributing the training points that are used to fit the volume fraction and the points to minimize the residual of the partial differential equations (PDE) is presented. Generally, it is found that 
the neural networks should be increased in width rather than length to increase their capability. Furthermore, it is shown that the weighting of the losses of the PDE residual is crucial for the 
optimizer to converge. Depending on the Weber number, the loss of the momentum equation residual can become orders of magnitude larger than all other losses due to the error between pressure gradient and surface tension term.
To compensate for this during the training procedure, the momentum equations must be weighted accordingly.
The Weber number of the test cases studied here range from $\mathcal{O}(-2)$ to $\mathcal{O}(1)$. When using the mean squared error as loss function, weighting the momentum equation residual with $\mathcal{O}(-3)$ to $\mathcal{O}(1)$, respectively, gives good results in 
regard to convergence of the loss associated with the volume fraction and the final loss of the momentum equation residual. The use of global and local (layer-wise) adaptive activation functions are evaluated. For all (inverse) test cases, the layer-wise adaptive activation functions are found
to consistently increase the convergence rate of the training process, whereas the application of global adaptive activation functions did not lead to any improvement. The improved training
speed is accompanied by an increase of the computational cost. For the large (inverse) cases investigated in this work, the average computational time per epoch increased by about 30\% compared to fixed activation functions.
However, there is no significant difference in computational cost between global and layer-wise adaptive activation functions.

\section*{Acknowledgements}
This project has received funding from the European Research Council (ERC) under the European Union's Horizon 2020 research and innovation programme (grant agreement No. 667483).
Furthermore, we would like to thank Prof. George Em Karniadakis of Brown University for stimulating discussions as part of a collaboration enabled by the Alexander von Humboldt foundation.

\appendix

\section{Comparison of the drop in a shear flow for different loss weights for the momentum equation residual and different activation functions}
\label{section:A}

\begin{figure}[!b]
    \centering
    \input{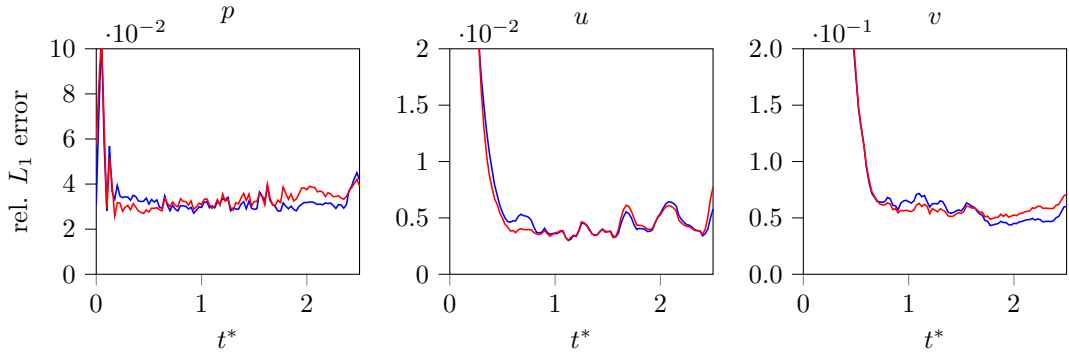}
    \caption{Comparison of the relative $L_1$ error between prediction and exact solution of the drop in a shear flow for different loss weights of the momentum equation residuum.
    Blue and red lines indicate loss weights of $w_{f,i}=\num{1e-3}$ and $w_{f,i}=\num{5e-3}$ with $i=\{u,v\}$, respectively.} 
    \label{fig:sheardrop_l1_loss_weights_variation}
\end{figure}

The weighting of the momentum equation residual is crucial for the training process. Here, loss weights $w_{f,i}=\num{1e-3}$ and $w_{f,i}=\num{5e-3}$ with $i=\{u,v\}$ are compared.
Note that all other loss weights are 1.0 and the hyper-parameters for training and neural network are the same as described in subsection \ref{subsection:sheardrop} 
In Table \ref{table:sheardrop_loss_weights_variation}, the final losses for both loss weights are listed. A reduction of the losses associated with the momentum equation residuum $MSE_{f,u}$ and $MSE_{f,v}$
of $-57\%$ and $-54\%$, respectively, is achieved. However, an increase of all other losses is observed. In particular, the loss associated with the volume fraction $MSE_\alpha$ increases by $194\%$. 
This corresponds to an increase of the mean interface thickness by $59\%$. The relative $L_1$ error between prediction and exact solution for both loss weights over time is 
depicted in Figure \ref{fig:sheardrop_l1_loss_weights_variation}. On average over time, there is no signficant difference with regard to the error between both models.
\begin{table}[!t]
    \centering
    \setlength\arrayrulewidth{1pt}
    \begin{tabular}{ccccccc}
        \hline
        Loss weights $w_{f,u/v}$ & $MSE_\alpha$ & $MSE_{BC}$ & $MSE_{f,m}$ & $MSE_{f,u}$ & $MSE_{f,v}$ & $MSE_{f,\alpha}$ \\
        \hline
        $\num{1e-3}$ & \num{1.22e-2} &\num{3.02e-4}&\num{4.13e-3}&\num{4.13e+0}&\num{3.49e+0}&\num{2.98e-3}\\
        $\num{5e-3}$ & \num{3.59e-2} &\num{5.38e-4}&\num{1.24e-2}&\num{1.79e+0}&\num{1.62e+0}&\num{4.05e-3}\\
        \hline
    \end{tabular}
    \caption{Final training loss of the drop in a shear flow for different loss weights for the momentum equation residuum.}
    \label{table:sheardrop_loss_weights_variation}  
\end{table}

\begin{figure}[!t]
    \centering
    \input{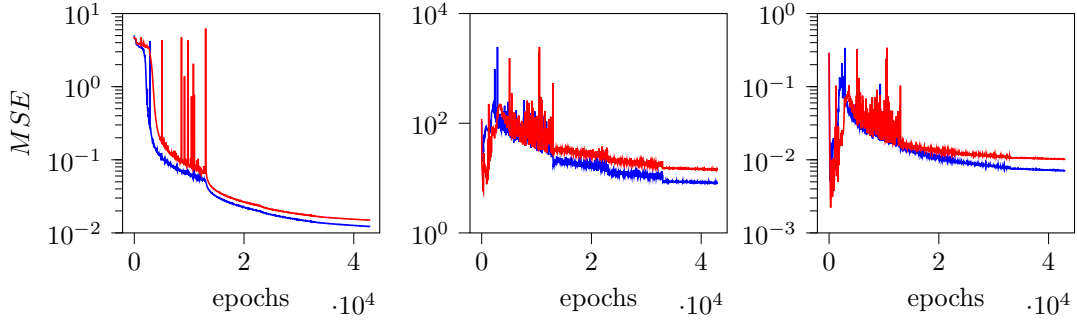}
    \caption{Training history of the drop in a shear flow for different activation functions. Blue and red lines indicate the hyperbolic tangent and sine function.
    (Left) Losses associated with the volume fraction $MSE_\alpha$. 
    (Middle) Losses associated with the momentum equation residuum $\sum_i MSE_{f,i}$, $i=\{u,v\}$.
    (Right) Losses associated with the residuum of the continuity equation and the advection equation $\sum_i MSE_{f,i}$, $i=\{m,\alpha\}$.} 
    \label{fig:sheardrop_loss_history_sin_tanh}
\end{figure}

Furthermore, we compare the hyperbolic tangent, which is used throughout this work, with a sine as activation function for the hidden layers. In Figure \ref{fig:sheardrop_loss_history_sin_tanh}, the loss history for both activation functions is illustrated. It shows that the sine activation leads to significantly more
noise in particular regarding the loss $MSE_\alpha$. Furthermore, the convergence rate of all losses is reduced.


\section{Comparison of the oscillating drop with and without velocity measurements on the boundaries}
\label{section:B}

\begin{figure}[!t]
    \centering
    \begin{tikzpicture}
        \node[anchor=south west] at (0,0) {\includegraphics[scale=0.5]{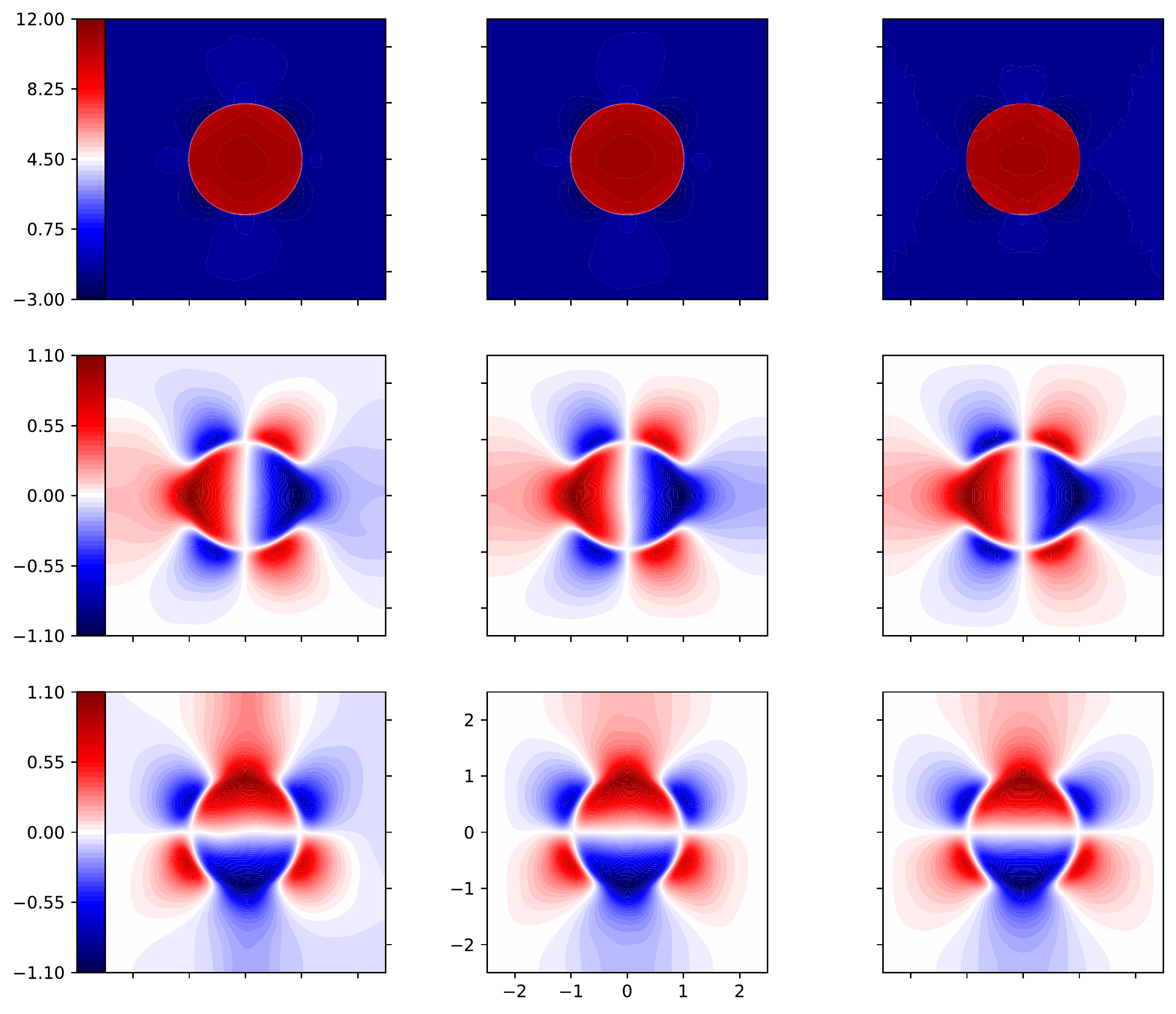}};
        \node at (6.6,0) {$x^*$};
        \node at (4.6,2.0) {$y^*$};
        \node[align=center, font=\linespread{1}\selectfont] at (2.5,10.8) {neural network \\ without measurements};
        \node[align=center, font=\linespread{1}\selectfont] at (6.6,10.8) {neural network \\ with measurements};
        \node[align=center, font=\linespread{1}\selectfont] at (10.6,10.8) {exact};
        \node[rotate=90, align=center, anchor=mid] at (-0.1,1.9) {$v^*$};
        \node[rotate=90, align=center, anchor=mid] at (-0.1,5.4) {$u^*$};
        \node[rotate=90, align=center, anchor=mid] at (-0.1,8.9) {$p^*$};
    \end{tikzpicture}
    \caption{Quantitative comparison of the prediction and exact solution of the oscillating drop at $t=0.33$ with and without velocitiy measurements
    at the boundaries.}
    \label{fig:oscillatingdrop_quantitative_ZG_F}
\end{figure}
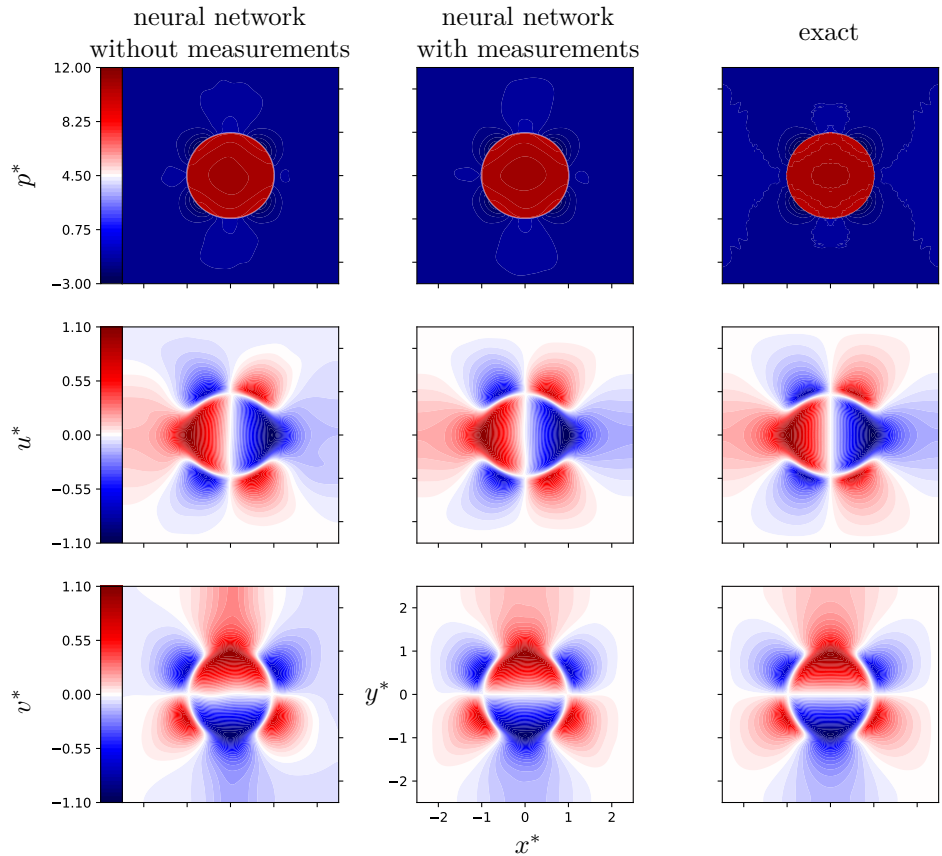

In subsection \ref{subsection:oscillatingdrop}, actual measurements of the velocity at the boundaries are given 
to the neural network to improve the prediction accuracy regarding the CFD solution. However, this is not necessary to infer the unique solution
that the CFD result provides. In Figure \ref{fig:oscillatingdrop_quantitative_ZG_F}, the predictions with and without measurements at the 
examplary time snapshot $t=0.33$ are compared to the CFD solution. Note that all training and neural network hyper-parameters are the same as described
in subsection \ref{subsection:oscillatingdrop}. It can be seen that the neural network predictions for the velocities close to and at the boundaries
are significantly worse when the neural network is not given actual measurements at the boundaries. This however may be improved by increasing the number of residual points within the domain, especially close to the boundaries, 
and by increasing the loss weights of the momentum equation residual.

\newpage

\bibliography{paper}

\end{document}